\documentclass[aps,longbibliography, prl,showpacs,twocolumn, superscriptaddress,amsmath,amssymb,verbatim]{revtex4-2}
 \usepackage[utf8]{inputenc}
 \usepackage{amsmath,amsfonts,amssymb}
 \usepackage{pifont}
 \usepackage{rotating}
 \usepackage{graphicx}    
 \usepackage{epstopdf}
 \usepackage{color}
 \usepackage[caption=false]{subfig}
 \usepackage{soul}
 \usepackage{booktabs}
 \usepackage{upgreek}
 \usepackage{hyperref}
 \definecolor{dark-red}{rgb}{0.9,0.15,0.15}
 \definecolor{dark-blue}{rgb}{0.15,0.15,0.4}
 \definecolor{medium-blue}{rgb}{0,0,0.5}
 \hypersetup{colorlinks=true,citecolor=blue,linkcolor=blue,urlcolor=blue}

\begin{document}
\title{ Novel Quantum Spin Liquid States in the $S = {\frac{1}{2}}$ Three-Dimensional Compound  Y$_{3}$Cu$_{2}$Sb$_{3}$O$_{14}$}
\author{Saikat Nandi}
\email{saikatnandi9@gmail.com}
\affiliation{Department of Physics, Indian Institute of Technology Bombay, Mumbai 400076, India}
\author{Rounak Das}
\affiliation{School of Physical Sciences, Indian Association for the Cultivation of Science, 2A and 2B Raja S.C. Mullick Road, Jadavpur, Kolkata 700 032, India}
\author{M. Hemmida}
\affiliation{Experimental Physics V, Center for Electronic Correlations and Magnetism, University of Augsburg, D-86135 Augsburg, Germany}
\author{M. U. Akbar}
\affiliation{Experimental Physics V, Center for Electronic Correlations and Magnetism, University of Augsburg, D-86135 Augsburg, Germany}
\author{H.-A. Krug von Nidda}
\affiliation{Experimental Physics V, Center for Electronic Correlations and Magnetism, University of Augsburg, D-86135 Augsburg, Germany}
\author{Jörg Sichelschmidt}
\affiliation{Max Planck Institute for Chemical Physics of Solids, 01187 Dresden, Germany}
\author{Sagar Mahapatra}
\affiliation{Department of Physics, Indian Institute of Science Education and Research, Pune, Maharashtra 411008, India}
\author{Marlis Schuller}
\affiliation{Experimental Physics V, Center for Electronic Correlations and Magnetism, University of Augsburg, D-86135 Augsburg, Germany}
\author{N. Büttgen}
\affiliation{Experimental Physics V, Center for Electronic Correlations and Magnetism, University of Augsburg, D-86135 Augsburg, Germany}
\author{J. M. Wilkinson}
\affiliation{ISIS Pulsed Neutron and Muon Source, STFC Rutherford Appleton Laboratory,
Harwell Campus, Didcot, Oxfordshire OX110QX, United Kingdom}
\author{M. P. Saravanan}
\affiliation{Low Temperature Laboratory, UGC-DAE Consortium for Scientific Research, Indore, 452001, India}
\author{Indra Dasgupta}
\affiliation{School of Physical Sciences, Indian Association for the Cultivation of Science, 2A and 2B Raja S.C. Mullick Road, Jadavpur, Kolkata 700 032, India}
\author{A.V. Mahajan}
\email{mahajan@phy.iitb.ac.in}
\affiliation{Department of Physics, Indian Institute of Technology Bombay, Mumbai 400076, India}

 \begin{abstract}
 
  The three-dimensional $S = {\frac{1}{2}}$ system Y$_{3}$Cu$_{2}$Sb$_{3}$O$_{14}$  consists of two inequivalent Cu$^{2+}$ sites, each forming an edge shared triangular lattice. Our magnetic susceptibility $\chi(T)$, specific heat $C_p(T)$,  $^{89}$Y nuclear magnetic resonance (NMR), muon spin relaxation ($\upmu\mathrm{SR}$), and electron spin resonance (ESR) measurements on this system confirm the absence of any long-range magnetic ordering and the persistence of spin dynamics down to 0.077 K. In $^{89}$Y NMR we find an anomaly at about 120 K which we suggest arises from a fraction of the spins condensing into a singlet (a valence bond solid VBS) state. A plateau in the muon relaxation rate is observed between 60 K and 10 K (signifying the VBS state from a fraction of the spins) followed by an increase and another plateau below about 1 K (presumably signifying the quantum spin liquid state from all the spins).   Our density functional theory calculations find a dominant antiferromagnetic interaction along the body diagonal with inequivalent  Cu(1) and Cu(2) ions alternately occupying the corners of the cube. All other near-neighbour interactions between the Cu ions are also found to be antiferromagnetic and are thought to drive the frustration.

\end{abstract}

\maketitle
\textit{Introduction$-$}   Quantum spin liquids (QSLs) are an exciting playground in many fields of physics, chemistry, theory of quantum computation, and materials science and engineering for exotic physical phenomena and emergent many-body quantum states. The starting point has often been spin-${\frac{1}{2}}$ systems with geometric frustration. This approach has led to the discovery of many systems such as the kagome material  ZnCu$_{3}$(OH)$_{6}$Cl$_{2}$ (or herbertsmithite), triangular-lattice organic compounds $\kappa$-(ET)$_2$Cu$_2$(CN)$_3$, and EtMe$_3$Sb[Pd(dmit)$_2$]$_2$, and vesignieite, BaCu$_{3}$V$_{2}$O$_{8}$(OH)$_{2}$, which show intriguing signs of QSL behavior~\cite{han2012fractionalized,okamoto2009vesignieite,zhou2017quantum}. 
  
       A variant of the above is a triangular lattice system LiZn$_{2}$Mo$_{3}$O$_{8}$ where, with decreasing temperature, the triangular magnetic lattice is found to decouple into a combination of free spins and a honeycomb lattice which then  condenses into a valence bond solid (VBS) state \cite{sheckelton2012possible,flint2013emergent,Sheckelton2014}.

   While most studies focus on quasi-two-dimensional systems, there has been some work on three-dimensional systems such as those with corner shared tetrahedra, hyperhoneycomb systems.  Three-dimensional (3D) arrangements of spin systems exhibit mostly ordered behavior in the absence of a special geometry.  QSL state in a $S = {\frac{1}{2}}$ system with a 3D geometry and  strong frustration has been realised in  PbCuTe$_{2}$O$_{6}$ ~\cite{chillal2020evidence}. 
   \begin{figure}[ht]
   	\begin{center}
   		\includegraphics[width=0.95\columnwidth]{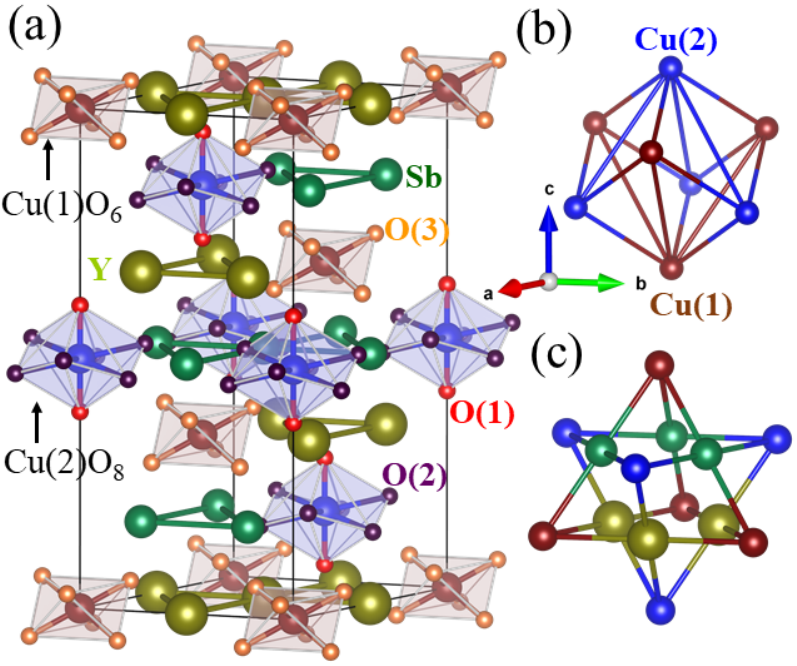}
   		\caption{(a) Crystal structure of Y$_{3}$Cu$_{2}$Sb$_{3}$O$_{14}$ in one unit cell consisting Cu(1)O$_6$  octahedra and Cu(2)O$_8$ hexagonal bipyramids. The atoms are shown in different colors. (b) Arrangement of the magnetic Cu$^{2+}$ forming 3D frustrated linkages. (c) Cu(1)-Y and Cu(2)-Sb layers composed of Cu(1) and Cu(2) networks.}
   		\label{F1}
   	\end{center}
   \end{figure}
     
     \begin{figure*}
     	\begin{center}
     		\includegraphics[width=2.05\columnwidth]{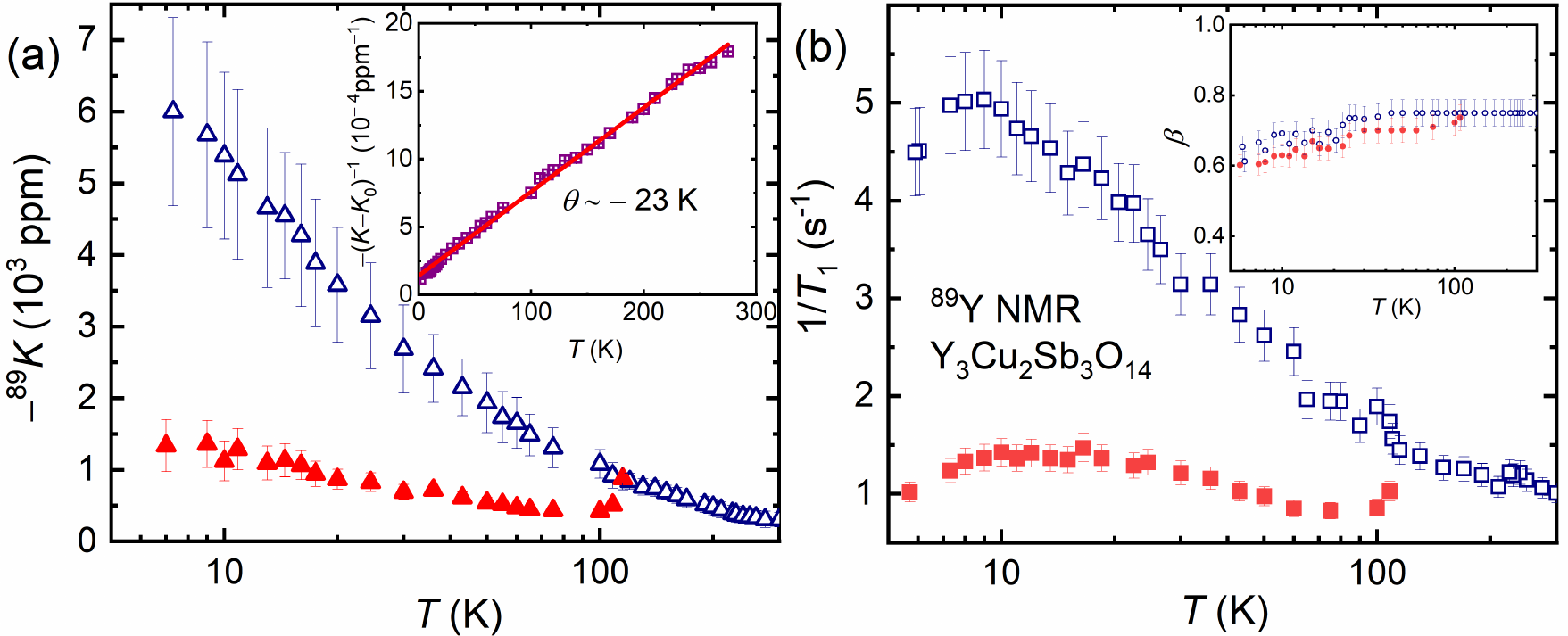}
     		\caption {(a) Variation of NMR shift, $^{89}K$ with temperature. Inset: (${K}$-${K}_0)^{-1}$ vs $T$ and solid red line represent Curie-Weiss fitting. Here, ${K}$ is the NMR shift of the more shifted line (open triangles in the main figure) (b) Spin-lattice relaxation rate ($1/T_1$) as a function of temperature.  The inset shows the stretching exponent $\beta$ as a function of temperature. }
     		\label{F2}	
     	\end{center}
     \end{figure*}
    \begin{figure}[h!]
    	\begin{center}
    		\includegraphics[width=0.95\columnwidth]{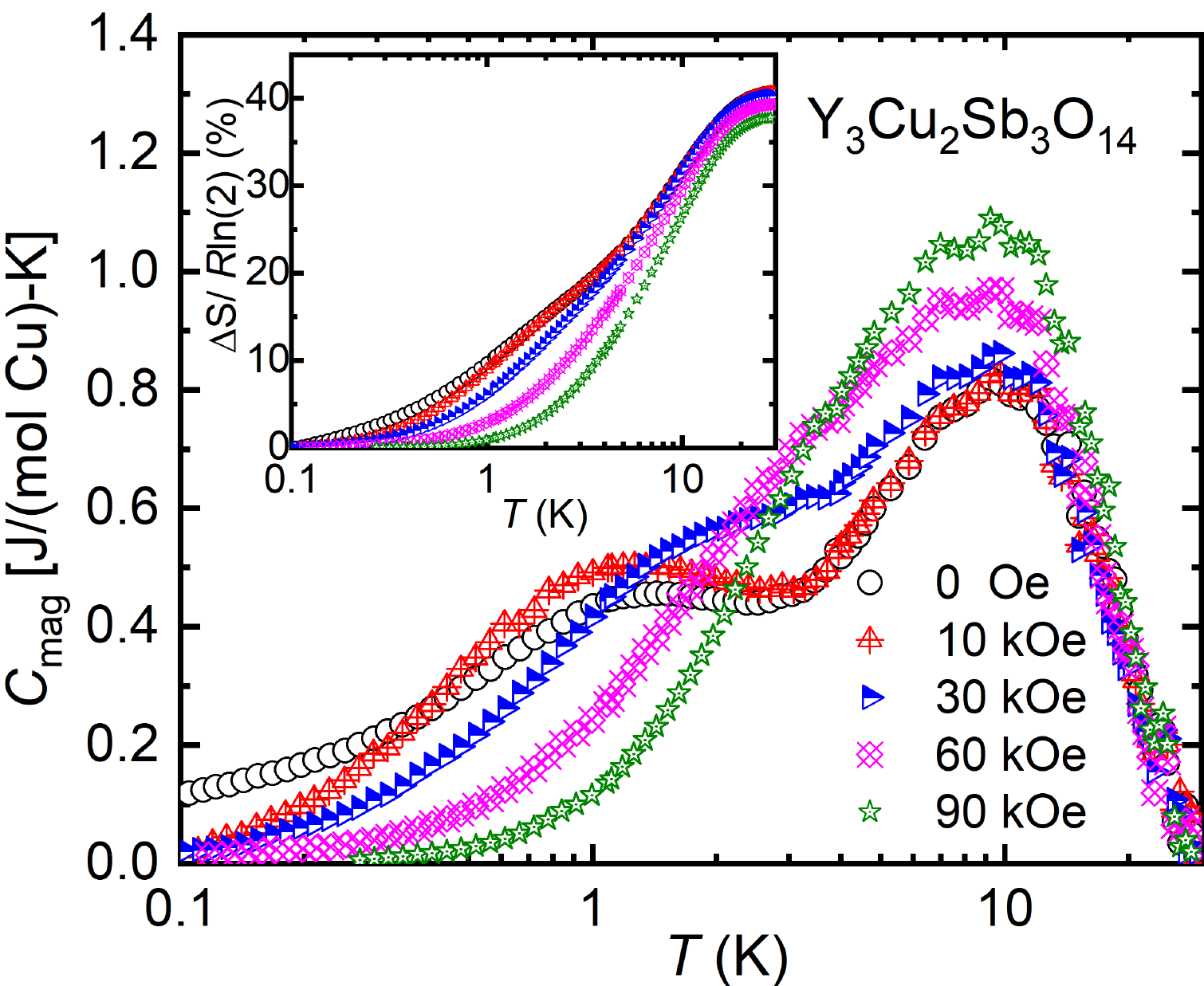}
    		\caption{The temperature dependence of the magnetic specific heat in several magnetic fields. The inset depicts the temperature dependence of associated magnetic entropy change $\Delta S$, normalized to $R$ln(2) per mol Cu.}
    		\label{F3}
    	\end{center}
    \end{figure}
    Our attention has been captured by  the rhombohedral pyrochlore family  (RE)$_{3}$M$_{2}$Sb$_{3}$O$_{14}$ (RE = Y, La, Pr, Nd, Sm, Eu, Gd, Tb, Dy, Ho, Er, Yb, Lu). This  series of materials has a rare-earth (RE) 2D kagome lattice separated by triangular planes of M$^{2+}$ ions~\cite{dun2016magnetic, dun2017structural, paddison2016emergent, fu2014crystal, kumar2022magnetic,lee2023random,lee2024random2,yang2021intrinsic,bradley2023high,li2014syntheses,mayouf2023synthesis,sanders2016aug,sanders2016re3sb3zn2o14,scheie2016}. However, the cubic phase is preferred with  decreasing size of the lanthanide ion and the compound Y$_{3}$Mn$_{2}$Sb$_{3}$O$_{14}$ has a cubic pyrochlore structure (space group $Fd\bar3m$) \cite{fu2014crystal}. 
    In the past few years, many exotic states have been  reported for (RE)$_{3}$Mg$_{2}$Sb$_{3}$O$_{14}$ (RE = Nd, Gd, Dy, Er) such as scalar chiral spin order, dipolar spin order, kagome spin ice,  Kosterlitz-Thouless transition \cite{dun2016magnetic}.  We have on the other hand focused on having a nonmagnetic ion at the rare-earth site while keeping a magnetic ion (Cu$^{2+}$ $S = 1/2$) on the M-site; i.e., Y$_{3}$Cu$_{2}$Sb$_{3}$O$_{14}$. The aim is to investigate for any novel states due to the triangular Cu$^{2+}$ lattice. 
    
     In this Letter, we report a comprehensive study of Y$_{3}$Cu$_{2}$Sb$_{3}$O$_{14}$ through bulk and local probe measurements.  While the absence of long-range order down to 0.077 K is evident, our $^{89}$Y nuclear magnetic resonance (NMR) data point to a fraction of the spins condensing into a VBS state below about 120 K followed by all the spins going into a gapless QSL state below about 1 K. A similar result is obtained from electron spin resonance (ESR) experiments where a part of the ESR signal intensity follows the Bleaney-Bowers dimer variation with temperature, with a broad maximum around 120 K (see Supplemental Material (SM) \cite{YCSO_Supp}~for details).  Muon spin relaxation measurements infer a muon depolarization rate ($\lambda_{\mathrm{ZF}}$) which increases with decreasing temperature and has plateaus below about 60 K  and again below 1 K.  The plateau at the higher temperature might signify a  decoupling of the frustrated lattice into two components: (i) an emergent lattice with fewer nearest neighbors for each lattice point which condenses into a VBS state and (ii) additional free spins. At the lower plateau in the muon relaxation rate, all the spins condense into a gapless QSL state. We suggest that ours is perhaps the first example of such a phenomenon in three dimensions and where the QSL state is reached through an intermediate VBS state.

        \textit{Experimental results$-$} Y$_{3}$Cu$_{2}$Sb$_{3}$O$_{14}$ crystallizes in the $R\bar{3}m$ space group at room temperature where Cu$^{2+}$ triangular layers are stacked by alternating Y and Sb kagome layers. Its crystal structure is illustrated in Fig. \ref{F1}(a).  There are two different crystallographic sites for Cu atoms corresponding to Cu(1) at the $3a$ or (0, 0, 0) and to Cu(2) at the $3b$ or (0, 0, 1/2) positions (see Ref. \cite{YCSO_Supp}~for details). 
      There is a 3-fold rotation symmetry around the $c$-axis at both the sites, as dictated by the crystal symmetry (see Fig. \ref{F1}(b, c)).

 The  $^{89}$Y NMR line splits below about 120 K (see Ref. \cite{YCSO_Supp}~for details) and the fractional line shift of the component which  continues to increase at lower-$T$ varies in a Curie-Weiss manner with an antiferromagnetic (AF) $\theta_{\rm CW} = -23(1)$ K (open symbols in Fig. \ref{F2}(a)). The other component (solid symbols in Fig. \ref{F2}(a)) has a shift $^{89}K(T)$ which drops sharply at about 120 K and remains small at lower temperatures consistent with condensation into a singlet state. Further, we have measured the $^{89}$Y  NMR  spin-lattice relaxation rate $T_1 ^{-1}$ (see Fig. \ref{F2}(b)) at each of the peaks to investigate the low-energy spin dynamics in Y$_{3}$Cu$_{2}$Sb$_{3}$O$_{14}$.   
The nuclear magnetization recovery is well fitted with a stretched exponential
{($M(t)=M_{0}[1-A\thinspace e^{-(t/T_{1})^{\beta}}]$)}.   At high temperature, $T_1 ^{-1}$ gradually increases as the temperature decreases. A sharp change is observed around 120 K, below which $T_1 ^{-1}$ is different at the two peak positions. $T_1 ^{-1}$ corresponding to the more shifted peak continues to increase with decreasing temperature (together with a Curie-like increase of its shift $K$) and finally decreases below 10 K. The $T_1 ^{-1}$ corresponding to the less shifted peak falls sharply (consistent with a line position that is nearly at the diamagnetic reference position) below 120 K, then increases slightly and exhibits a plateau from 20 K to 10 K and drops thereafter.
A broad maximum in $T_1 ^{-1}$ has been observed in numerous organic spin liquid candidates \cite{Pustogow2020, Pustogow2022, Itou2023}, where it is attributed to the relaxation of impurity spins, and this might be the case here as well.
The decrease of the effective moment below about 120 K is demonstrated by having a fraction of the spins dropping out from the CW behavior and becoming a part of the VBS phase.  
 It is reasonable to say that the slower component arises from the fraction that has condensed into a VBS state while the faster component is from the fraction that remains paramagnetic below 120 K but eventually also condenses into a QSL state below about 10 K. In lab-based XRD, we did not observe any change in the structure around 120 K as one might expect from a symmetry breaking. However, it is possible that there is only a subtle structural change (a small change in bond lengths and bond angles) associated with the symmetry breaking VBS transition. This  partial VBS formation could also be due to chemical disorder.
 This scenario is akin to that in LiZn$_2$Mo$_3$O$_8$ where the triangular lattice decouples into a honeycomb lattice (which condenses into a VBS state) plus free spins at the center of the honeycombs \cite{sheckelton2012possible,flint2013emergent,Sheckelton2014}. In the present case, given the three-dimensional geometry, it is not obvious what the decoupled lattice might be though it is primarily thought to be made of body-diagonals of the distorted cubes in the structure (see Fig. \ref{F1} (b)).

We next examine the low-energy excitations and spin dynamics through specific heat and muon spin relaxation ($\upmu\mathrm{SR}$) measurements.
The magnetic specific heat $C_\mathrm{mag} (T)$ (see Fig. \ref{F3}) was obtained by subtracting the lattice contribution obtained from the data of the nonmagnetic structural analog Lu$_3$Zn$_2$Sb$_3$O$_{14}$ (see Ref. \cite{YCSO_Supp}~for details).  $C_\mathrm{mag} (T)$ exhibits a broad hump at 10 K followed by a plateau around 1 K. The low-$T$ plateau is suppressed in fields greater than 30 kOe. The entropy change calculated by integrating the $C_\mathrm{mag}(T)/T$ vs. $T$ data is shown in Fig. \ref{F3} (inset). Close to 40\% of the entropy (compared to $R\ln2$) is recovered by about 20 K. In our suggested scenario, a fraction of the spins condense into  a VBS below about 120 K, while all the spins condense to a gapless QSL state below about 10 K. So, a part of the entropy release would happen at about 120 K while the remaining would be around 10 K. We observe the lower temperature plateau around 20 K while the higher temperature plateau is difficult to obtain reliably due to the uncertainty in inferring the magnetic contribution by subtracting the lattice contribution which is large and dominant already beyond 20 K. 

\begin{figure}[ht]
	\begin{center}
		\includegraphics[width=0.95\columnwidth]{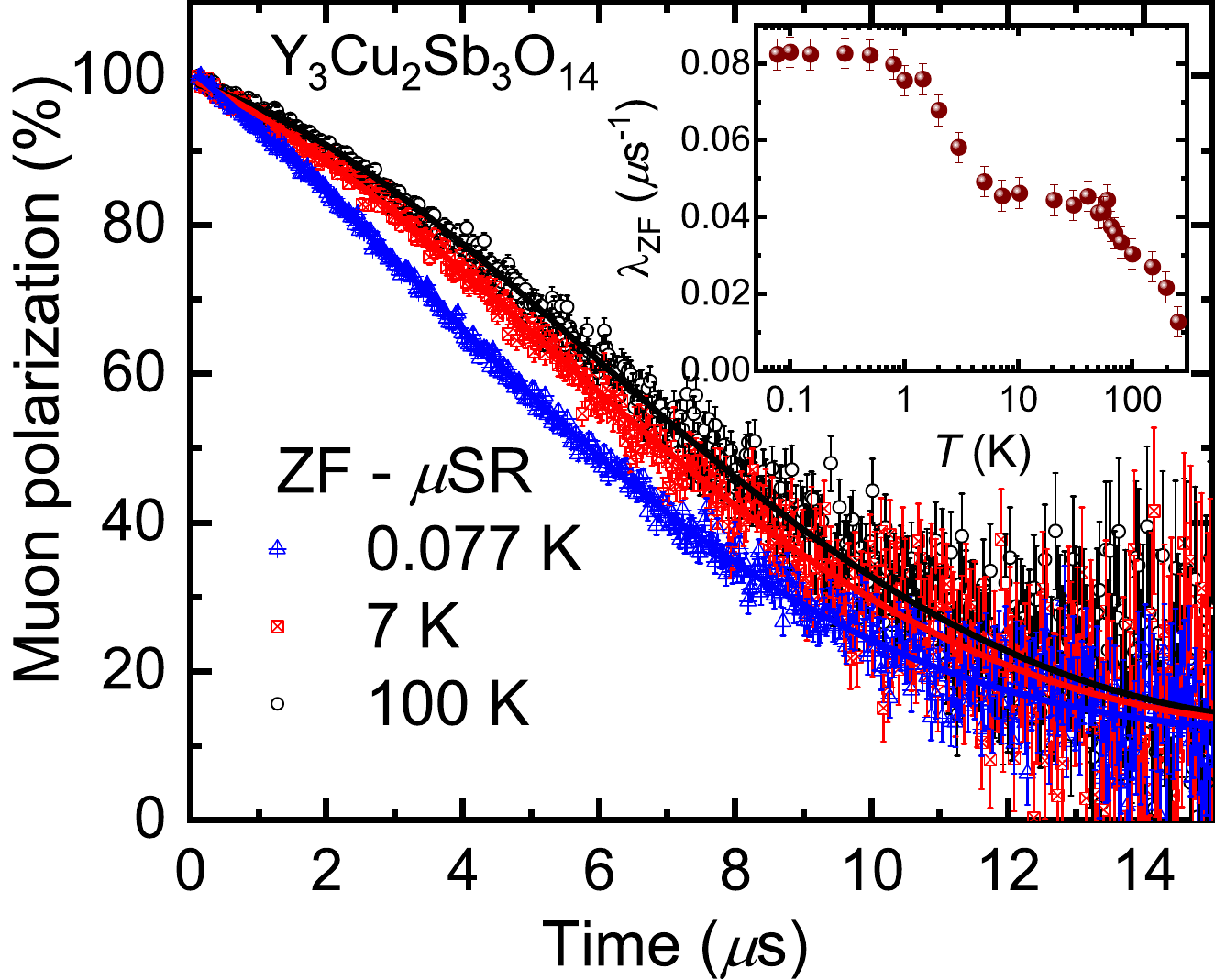}
		\caption{The time evolution of the ZF-$\upmu\mathrm{SR}$ asymmetry spectra at selected temperatures where solid lines are the fits to Eq. (\ref{eq1}). Inset shows the temperature dependence of the muon spin relaxation rate, $\lambda_{\mathrm{ZF}}$ extracted from fits to the data.}
		\label{F4}
	\end{center}
\end{figure}

We next detail our $\upmu\mathrm{SR}$ data and analysis which show signatures of two plateaus with decreasing temperature.
 The time-dependent zero-field (ZF) asymmetry  spectra for Y$_{3}$Cu$_{2}$Sb$_{3}$O$_{14}$  displayed in Fig. \ref{F4} confirm the absence of any coherent oscillations or any initial asymmetry drop in the spectra down to 0.077 K. Furthermore, there is no indication of a  1/3-recovery of the muon polarization at long times which precludes the occurrence of any long-range static magnetic order. The ZF $\upmu\mathrm{SR}$ spectra are well fitted by a Gaussian Kubo-Toyabe (KT) function, $G_{ZF}^{KT},$ multiplied by an exponential function, and a temperature-independent background, $P_{BG}$ arising from the silver sample holder or the cryostat:
\begin{equation} \label{eq1}
P(t) = P_{0} \exp(-\lambda_{\mathrm{ZF}}t) G_{ZF}^{KT} (\Delta, t) + P_{BG}
\end{equation}
 where, $P_{0}$ is the  initial asymmetry and $\lambda_{\mathrm{ZF}}$ represents the ZF relaxation rate caused by the electronic Cu spins (see Ref. \cite{YCSO_Supp}~for details).
 
The obtained $\lambda_{\mathrm{ZF}}$ is plotted against temperature in Fig. \ref{F4}(inset).  With decreasing temperature $\lambda_{\mathrm{ZF}}$ increases and then holds a plateau in the temperature range 65 K $\gtrsim T \gtrsim$ 10 K. With further cooling below 10 K, $\lambda_{\mathrm{ZF}}$ increases gradually, and then, saturates at about $T \thickapprox$ 1 K and remains constant down to the lowest measured temperature 0.077 K. Our longitudinal field data (see Ref. \cite{YCSO_Supp}~for details) show that the muon spin is not decoupled from the internal moments even in fields of 3200 Oe. This result shows that Y$_{3}$Cu$_{2}$Sb$_{3}$O$_{14}$ enters a low-$T$ regime where the spin dynamics are persistent and temperature independent. Such behavior has been observed in numerous other proposed QSL candidates \cite{Kundu2020, choi2019exotic}.
Finally, the bulk  magnetic susceptibility  $\chi (T)$ (see Ref. \cite{YCSO_Supp}~for details) varies in a Curie-Weiss manner (high-temperature Curie-Weiss temperature $\theta_{\rm CW} \simeq - 23 (1)$ K and an effective moment ${\mu}_{\mathrm{eff}}=1.57{\mu}_B$), with the absence of any anomalies suggesting the absence of any long-range order down to 0.40 K.  Furthermore, no apparent difference is seen between the zero field cooled (ZFC) and field cooled (FC) susceptibility $\chi(T)$ data taken in an applied magnetic field $H \approx 100$ Oe which excludes the occurrence of spin freezing. Also, from a plot of the inverse susceptibility with temperature (see Ref. \cite{YCSO_Supp}~for details), we find a change of slope around 60 K below which the Curie term shows a small decrease. 
The ESR data allow to identify the fraction of copper spins forming dimers below 120\,K as about 20\% of all copper spins. Its intensity can be well described by the characteristic Bleaney-Bowers law for the magnetic susceptibility of spin dimers (see Ref. \cite{YCSO_Supp}~for details). The remaining 80\% are ESR silent probably due to strong spin-phonon interaction like in the high-$T_{c}$ cuprates. Only a small fraction of these unpaired spins shows up in the ESR spectra at low temperature, revealing a very large linewidth which strongly increases with increasing temperature, so that the signal becomes undetectable above 25\,K. This supports the assumption of randomly distributed antiferromagnetic couplings between the residual spins after dimerization of the 20\% fraction. 


\textit{Electronic structure calculation$-$}
To check the viability of a quantum spin liquid, we carried out first principles electronic structure calculations based on density functional theory (DFT) for the experimentally determined rhombohedral pyrochlore structure of Y$_3$Cu$_2$Sb$_3$O$_{14}$. All the electronic structure calculations were carried out using DFT in the pseudopotential plane-wave basis within the generalized gradient approximation \cite{PBE1996} supplemented with Hubbard $U$ \cite{HubbardU1991} as encoded in the Vienna $ab$ $initio$ simulation package \cite{abinitio1993, abinitio1996} with projector augmented wave potentials \cite{PAW1994, pseudopotential1999}. The calculations were done with standard values of $U_{\mathrm{eff}} {\equiv} U_d-J_H=$ 6.5 eV \cite{HubbardU1991} chosen for Cu. The non-spin polarized total, Cu-$d$ and O-$p$ partial density of states for Y$_3$Cu$_2$Sb$_3$O$_{14}$ (see Ref. \cite{YCSO_Supp}~for details) reveal that the Fermi level is dominated by half filled Cu(1) $e_g$ and Cu(2) $a_{1g}$ states and the O-$p$ states are fully occupied while the Sb-$s$, Sb-$p$, Y-$s$ and Y-$d$ states are completely empty consistent with the nominal ionic formula of this compound, Y$_3$$^{3+}$Cu$_2$$^{2+}$Sb$_3$$^{5+}$O$_{14}$$^{2-}$ (see Ref. \cite{YCSO_Supp}~for details). Spin polarized calculations with ferromagnetic arrangement of Cu spins yields a total moment of 2.0 ${\mu}_B$ per formula unit, which further supports the $S = \frac{1}{2}$ moment of the Cu atom and is consistent with the experimentally determined effective moment (${\mu}_{\mathrm{eff}}=1.57{\mu}_B)$. However, the calculated magnetic moment per Cu site is 0.80 ${\mu}_B$ (0.80 ${\mu}_B$ for Cu(1) and 0.76 ${\mu}_B$ for Cu(2)), where the rest of the moment is shared with the ligand atoms due to substantial hybridization of the Cu atoms with ligands.

\begin{figure}[ht]
	\begin{center}
		\includegraphics[width=1.0\columnwidth]{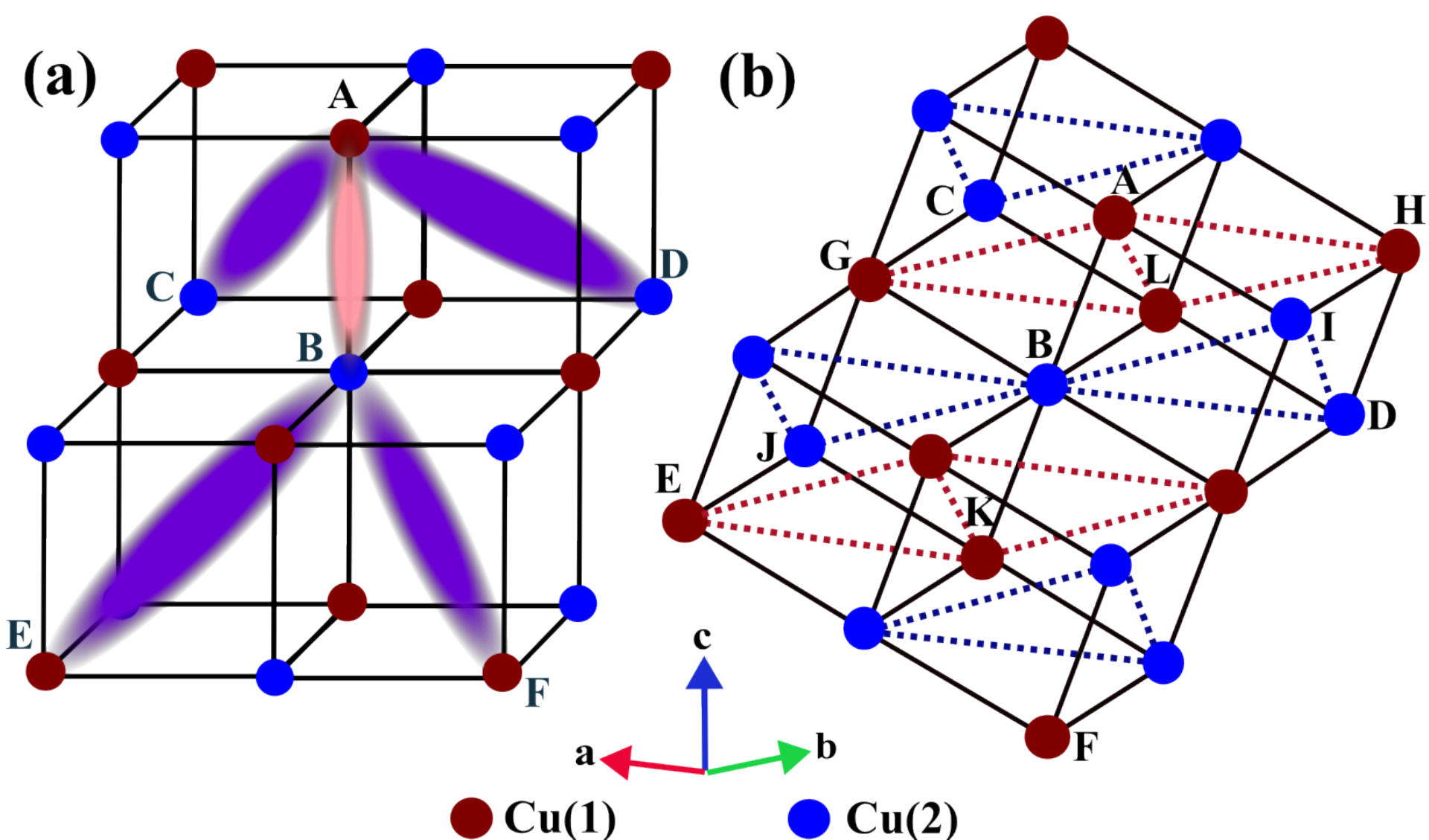}
		\caption{(a) A schematic of the slightly distorted cubic structure formed by the nearest-neighbor Cu-Cu network in Y$_3$Cu$_2$Sb$_3$O$_{14}$. The brown dots represent Cu(1) atoms and blue dots represent Cu(2) atoms. The different exchange paths are also shown. (b) The different in-plane triangular networks formed by the Cu atoms are shown by the dotted lines}
		\label{F5}
	\end{center}
\end{figure}
The crystal structure is such that, the nearest-neighbor arrangement between Cu(1) and Cu(2) forms a slightly distorted cubic structure with unequal body-diagonals as shown in Fig. \ref{F5}, where each Cu(2) is a nearest-neighbor to a Cu(1). The exchange path between two inequivalent Cu atoms forming the body diagonal is shown in Fig. \ref{F5}(a) with violet shades. In addition, we have shown the exchange path along the face diagonals either between Cu(1)and Cu(1) or Cu(2)and Cu(2) shown in Fig. \ref{F5}(b), forming a triangular network. 

We have calculated these exchange interactions by employing the  \lq{four state}\rq~method \cite{fourstate2011, fourstate2019, fourstate2021}. This is based on a computation of the total energy of the system with collinear spin alignment, where the spin configurations on the two chosen sites are modified while restricting the rest of the spins to a ferromagnetic configuration. The calculated exchange parameters along body diagonals are presented in Table \ref{tab:DFT}. The exchange parameters along the face diagonal where each of the Cu(1) and Cu(2) ions form a triangular network, are shown in Table \ref{tab:DFT_1}. Calculations reveal that the dominant exchange interactions are along the body diagonals of the cube, forming dimers where the inter-dimer exchange interaction ($J_{AB}$) is negligible compared to the intradimer exchange interactions. The difference between the intradimer and inter-dimer interaction may be attributed to the geometry of the super-super exchange path mediated by oxygens which is nearly linear for the intradimer exchange and substantially distorted for the inter-dimer exchange path. As a consequence, in spite of the short interdimer distance, the intradimer exchange interaction dominates \cite{Walker2016}. This opens up the possibility for the formation of VBS within our system. The antiferromagnetic interaction along the face diagonal of the cube forming a triangular network for Cu(1) and Cu(2) introduces frustration and prohibits long-range ordering. Our first principles DFT results corroborate well with the experimental results, and lend credence to the scenario of spin liquid behavior in this three-dimensional system induced by frustration.
\begin{table}[h]
	\centering
	\caption{Exchange parameters along the body diagonals.} \label{tab:DFT}
	\renewcommand{\arraystretch}{1.2} 
	\setlength{\tabcolsep}{4pt} 
	\resizebox{\columnwidth}{!}{ 
		\begin{tabular}{c c c}
			\toprule
			$J$ & Cu-Cu distance (\AA) & Exchange parameter (meV) \\
			\midrule
			$J_{AC}$   & 9.01 & 3.6 \\
			$J_{AD}$ & 9.01 & 4.6 \\
			$J_{BE}$ & 9.01 & 3.6 \\
			$J_{BF}$ & 8.58 & 3.0 \\
			$J_{AB}$ & 5.14 & 0.2 \\
			\bottomrule
		\end{tabular}
	} 
	\label{table:hopping}
\end{table}
\begin{table}[h]
	\centering
	\caption{Exchange parameters along the face diagonals or in the Cu(1) planes and Cu(2)  planes.} \label{tab:DFT_1}
	\renewcommand{\arraystretch}{1.2} 
	\setlength{\tabcolsep}{4pt} 
	\resizebox{\columnwidth}{!}{ 
		\begin{tabular}{c c c}
			\toprule
			$J$ & Cu-Cu distance (\AA) & Exchange parameter (meV) \\
			\midrule
			$J_{AH}$ & 7.40 & 2.2 \\
			$J_{AL}$ & 7.40 & 1.6 \\
			$J_{HL}$ & 7.40 & 1.5 \\
			$J_{BI}$ & 7.40 & 2.0 \\
			$J_{ID}$ & 7.40 & 2.1\\
			$J_{BD}$ & 7.40 & 1.7 \\
			\bottomrule
		\end{tabular}
	} 
	\label{table:hopping}
\end{table}

\textit{Conclusion$-$} Our comprehensive measurements on Y$_{3}$Cu$_{2}$Sb$_{3}$O$_{14}$ have established that in spite of a sizable exchange interaction corresponding to an AF $\theta_{\rm CW} = -23$ K, the compound does not order down to 0.077 K. More importantly, unlike other candidate QSLs, the present compound shows two crossovers with decreasing temperature. The signatures of the higher temperature condensation (bifurcation at 120 K in NMR and a plateau below about 60 K in  $\upmu\mathrm{SR}$ $\lambda_\mathrm{ZF}$, and a dimer fit of the ESR intensity) are apparent in our  data, while those of the low-$T$ one are distinctly seen in our $\upmu\mathrm{SR}$ and ESR data though hints are also present in NMR and specific heat. 

 Backed by the evidence we have presented, we suggest that the higher temperature plateau  corresponds to the condensation of a fraction of the spins into a VBS state, while all the spins condense into a QSL state below 1 K.
 This novel result in a three-dimensional frustrated magnet should spark intense activity to examine magnetic excitations through neutron scattering in this compound.

\textit{Acknowledgments$-$} 
We thank DST-SERB for funding under the DST-CRG scheme (CRG/2021/003024). We acknowledge the use of various central Facilities at IIT Bombay. Experiments at the ISIS Neutron and Muon Source
were supported by a beam-time allocation RB2410186 \cite{RAL} from the Science
and Technology Facilities Council UK (funded via the ISIS-Indian Partnership agreement). We acknowledge UGC-DAE CSR (Indore) for providing the mK specific heat experimental facility (01/2024/6185). We thank Dr. Rahul Kumar and Dr. A. Sundaresan (Jawaharlal Nehru Centre for Advanced Scientific Research) for their support with dc magnetization measurements using a $^3$He attachment to the MPMS. I. D. and R. D. would like to thank Dr. Atasi Chakraborty for discussions. S.N. would like to acknowledge funding support for the Chanakya Postdoctoral fellowship (CPDF/2021-22/01) from the National Mission on Interdisciplinary Cyber Physical Systems, of the DST, Govt. of India through the I-HUB Quantum Technology Foundation. This work was
partially supported by the Deutsche Forschungsgemeinschaft
(DFG) within the Transregional Collaborative Research
Center TRR 360 “Constrained Quantum Matter”,
Project No. 492547816 (Augsburg, Munich, Stuttgart,
Leipzig).
\bibliography{ref}

\end{document}


\title{ Supplemental Material: Novel Quantum Spin Liquid States in the $S = {\frac{1}{2}}$ Three-Dimensional Compound  Y$_{3}$Cu$_{2}$Sb$_{3}$O$_{14}$}
\author{Saikat Nandi}
\email{saikatnandi9@gmail.com}
\affiliation{Department of Physics, Indian Institute of Technology Bombay, Mumbai 400076, India}
\author{Rounak Das}
\affiliation{School of Physical Sciences, Indian Association for the Cultivation of Science, 2A and 2B Raja S.C. Mullick Road, Jadavpur, Kolkata 700 032, India}
\author{M. Hemmida}
\affiliation{Experimental Physics V, Center for Electronic Correlations and Magnetism, University of Augsburg, D-86135 Augsburg, Germany}
\author{M. U. Akbar}
\affiliation{Experimental Physics V, Center for Electronic Correlations and Magnetism, University of Augsburg, D-86135 Augsburg, Germany}
\author{H.-A. Krug von Nidda}
\affiliation{Experimental Physics V, Center for Electronic Correlations and Magnetism, University of Augsburg, D-86135 Augsburg, Germany}
\author{Jörg Sichelschmidt}
\affiliation{Max Planck Institute for Chemical Physics of Solids, 01187 Dresden, Germany}
\author{Sagar Mahapatra}
\affiliation{Department of Physics, Indian Institute of Science Education and Research, Pune, Maharashtra 411008, India}
\author{Marlis Schuller}
\affiliation{Experimental Physics V, Center for Electronic Correlations and Magnetism, University of Augsburg, D-86135 Augsburg, Germany}
\author{N. Büttgen}
\affiliation{Experimental Physics V, Center for Electronic Correlations and Magnetism, University of Augsburg, D-86135 Augsburg, Germany}	
\author{J. M. Wilkinson}
\affiliation{ISIS Pulsed Neutron and Muon Source, STFC Rutherford Appleton Laboratory,
	Harwell Campus, Didcot, Oxfordshire OX110QX, United Kingdom}
\author{M. P. Saravanan}
\affiliation{Low Temperature Laboratory, UGC-DAE Consortium for Scientific Research, Indore, 452001, India}
\author{Indra Dasgupta}
\affiliation{School of Physical Sciences, Indian Association for the Cultivation of Science, 2A and 2B Raja S.C. Mullick Road, Jadavpur, Kolkata 700 032, India}
\author{A.V. Mahajan}
\email{mahajan@phy.iitb.ac.in}
\affiliation{Department of Physics, Indian Institute of Technology Bombay, Mumbai 400076, India}

\maketitle

  Herein we report details of various measurements together with \textit{ab initio} electronic structure calculations performed by us on Y$_{3}$Cu$_{2}$Sb$_{3}$O$_{14}$.

\subsection {Experimental details}
 Polycrystalline Y$_{3}$Cu$_{2}$Sb$_{3}$O$_{14}$ samples were prepared by conventional solid state reaction techniques using high purity starting materials. Y$_{2}$O$_{3}$, Sb$_{2}$O$_{3}$, CuO were thoroughly ground with a mortar-pestle and the homogeneous powder was pressed into disk-shaped pellets. The samples were then subject to heat treatments in air using a box furnace at ${900}^\circ $C and ${1100}^\circ$C for 12 h and 24 h, respectively, with intermediate grinding and pelletization. Finally, we sintered the samples at ${1200}^\circ$C in air. To determine the crystal structure, powder X-ray diffraction (XRD) measurement were performed at room temperature with a high resolution Rigaku diffractometer using Cu-K$_\alpha$ radiation ($\lambda$ = 1.5406 Å) and structural refinement was performed through the Rietveld method as implemented in FULLPROF~\cite{rodriguez1993recent}.   The dc and ac magnetization ($M$) measurements were performed in the temperature range 1.8–350 K and in various applied magnetic fields (0 Oe $\leq H\leq 70 $ kOe), with a Quantum Design SQUID Vibrating Sample Magnetometer (SVSM). In addition, dc magnetization measurements were performed in the temperature range 0.40 K $\leq T\leq 1.80 $ K using a $^3$He (iQHelium3, Quantum Design) attachment to the
 magnetic properties measurement system (MPMS). For ac magnetic susceptibility measurements, we used an ac drive field of 3 Oe at frequencies $\nu = 11 - 941$ Hz under no external dc field. We also measured field and temperature dependent ac susceptibility under various applied dc magnetic fields $H =$ 100 Oe, 1 kOe, 10 kOe, 20 kOe, 30 kOe at a fixed frequency of $\nu= 111$ Hz. Specific heat ($C_\mathrm{p}$) as a function of $T$ (2 K $\leq T\leq 250 $ K) and $H$ (0 Oe $\leq H\leq 90 $ kOe) was measured using a thermal relaxation technique in a physical property measurement system (PPMS, Evercool-II, Quantum Design). In addition, low-temperature specific heat measurements were carried out down to 0.10 K under magnetic fields 0 Oe $\leq H \leq 90 $ kOe  using a dilution refrigerator insert in a PPMS Dynacool (Quantum Design).  $^{89}$Y nuclear magnetic resonance (NMR) measurements have been done using pulsed NMR techniques in a fixed field of 93.9543 kOe with a Tecmag spectrometer. The temperature was varied  with an Oxford continuous-flow cryostat. NMR Measurements have also been performed with a cryofree magnet (Cryomagnetics Inc.) in the field-sweep mode at a fixed frequency of 16.69 MHz down to 6 K. From our measurements, we obtained $^{89}$Y NMR spectra, spin-lattice relaxation rates ($1/T_1$) as a function of temperature.  
 We used a continuous wave electron spin resonance (ESR) spectrometer (Bruker Elexsys E500A) working at X-band frequencies ($\nu = 9.4$ GHz). The powder sample was embedded in paraffin and the temperature varied inthe range 4 K $\leq T\leq 300$ K using a helium-flow cryostat. Muon spin relaxation ($\upmu\mathrm{SR}$) measurements were carried out using the MUSR spectrometer at the ISIS Neutron and Muon Source at the STFC Rutherford Appleton Laboratory in the UK. The powder sample (thickness $\sim$2 mm) was loaded onto a pure silver plate (99.995+$\%$) using GE varnish and was covered with a thin (18 $\mu$m) silver foil. The spectra were collected at elevated temperatures between 5 K and 200 K in a $^{4}$He cryostat, and at low temperatures, down to 0.077 K in a dilution refrigerator. All of the obtained ZF and LF $\upmu\mathrm{SR}$ data were analyzed using the WiMDA software package \cite{Pratt2000}.  
\begin{table}[h]
	\centering{}\caption{\label{tab:Rietveld-parameter_YCSO}{Obtained lattice parameters and the goodness-of-fit parameters after the Rietveld refinement.}}
	\begin{tabular}{|lll|}
		
		\hline 
		Space group & $R{\bar 3}m$ & \tabularnewline
		Lattice parameter  & $\mathit{a}$ = $\mathit{b}$ = 7.399(0) \AA & \tabularnewline
		& $\mathit{c}$ = 17.168(0) \AA & \tabularnewline
		$\alpha$, $\beta$, $\gamma$ & $\alpha=\beta=  $90$^{\mathrm{o}}, \gamma=$120$^{\mathrm{o}}$ & \tabularnewline
		Cell volume {[}\AA $^{3}${]} & 813.969 &  \tabularnewline
		$R_{\rm p}$, $R_{\rm wp}$, $R_{\rm exp}$(\%) & 22.1, 19.4, 15.2 &  \tabularnewline
		Bragg $R$-factor(\%) & 8.52 & \tabularnewline
		RF-factor(\%) & 8.51 &  \tabularnewline
		$\chi^{2}$ & 1.613 & \tabularnewline
		\hline 
	\end{tabular}
\end{table}
\begin{table}[h]
	\centering{}\caption{\label{tab:Atomic-positions-in YCSO}{Atomic coordinates with their respective site occupancy and isotropic Debye-Waller factors ($B_{iso}$) in the unit cell of 
			Y$_{3}$Cu$_{2}$Sb$_{3}$O$_{14}$ based on a $R\bar{3}m$ space group refinement.}}
	\vspace{0.5cm}
	\begin{tabular}{cccccc}
		\hline 
		Atom & $x/a$ & $y/b$ & $z/c$ & Occupancy & $B_{iso}$(\AA$^2)$ \tabularnewline
		(Wyckoff position)
		\tabularnewline
		\hline 
		\hline 
		Y~~~~~~ ($9e$) & 0.500 & 0.000 & 0.000 & 1.000 & 0.340\tabularnewline
		Cu(1)~ ($3a$) & 0.000 & 0.000 & 0.000 & 1.000 & 0.190\tabularnewline
		Cu(2)~ ($3b$) & 0.000 & 0.000 & 0.500 & 1.000 & 0.040\tabularnewline
		Sb~~~~~ ($9d$) & 0.500 & 0.000 & 0.500 & 1.000 & 0.670\tabularnewline
		O(1)~~  ($6c$) & 0.000 & 0.000 & 0.394 & 1.000 & 0.550\tabularnewline
		O(2) ($18h$) & 0.510 & 0.490 & 0.143 & 1.000 & 0.710\tabularnewline
		O(3) ($18h$) & 0.137 & 0.863 & -0.049 & 1.000 & 0.800\tabularnewline
		\hline 
	\end{tabular}
\end{table}
\renewcommand{\thefigure}{S\arabic{figure}}  
\setcounter{figure}{0} 
\begin{figure}[ht]
	\begin{center}
		\includegraphics[width=0.95\columnwidth]{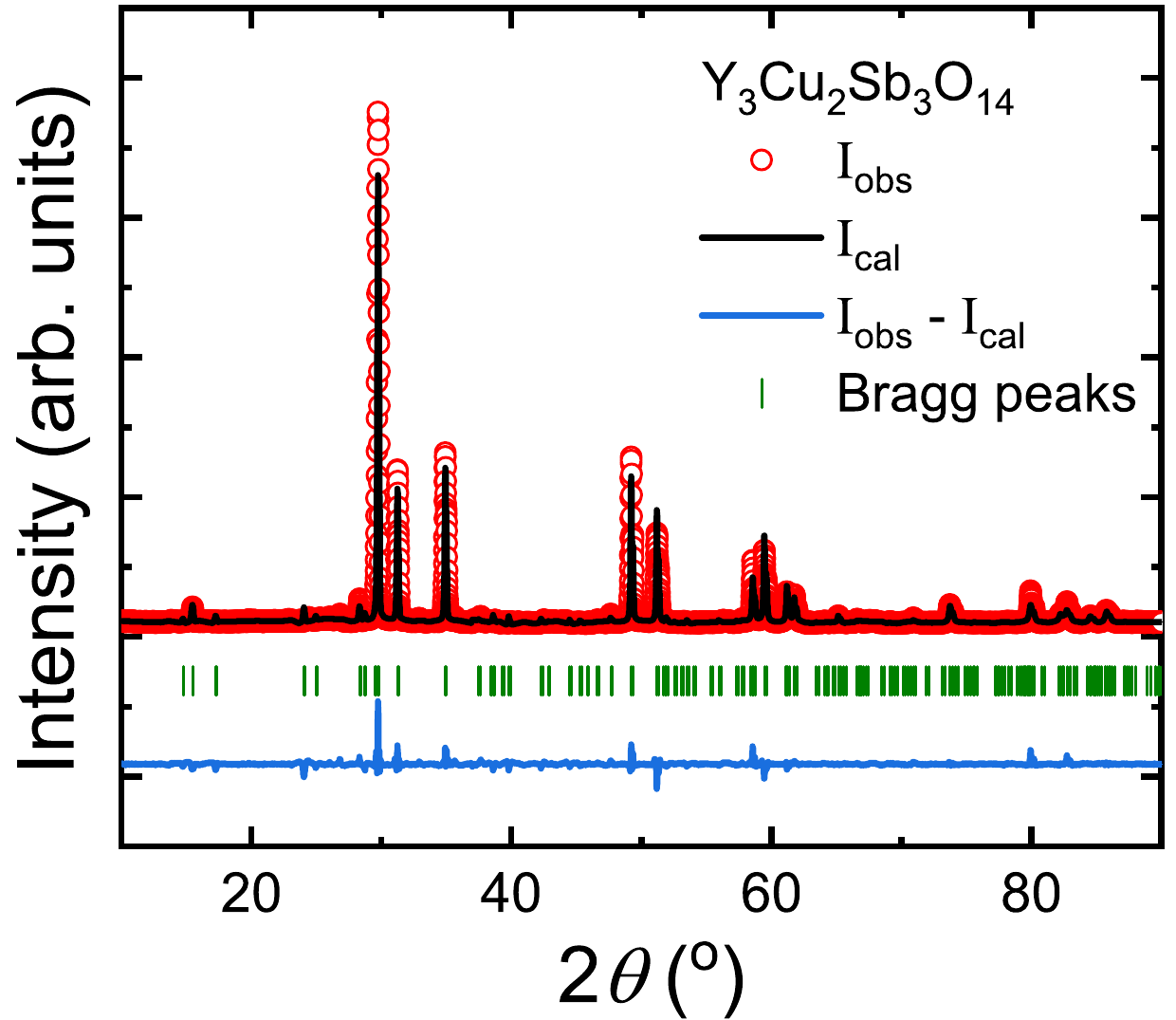}
		\caption{Rietveld refinement of the powder XRD pattern for Y$_{3}$Cu$_{2}$Sb$_{3}$O$_{14}$ at room temperature. The red circles represent the observed (I$_\mathrm{obs}$) intensity, whereas the calculated (I$_\mathrm{cal}$) patterns and the difference (I$_\mathrm{obs}-$I$_\mathrm{cal}$) are shown in black and blue lines, respectively. The green tick marks denote the allowed Bragg positions.}
		\label{fig:S1}
	\end{center}
\end{figure}
\begin{figure}[ht]
	\begin{center}
		\includegraphics[width=0.95\columnwidth]{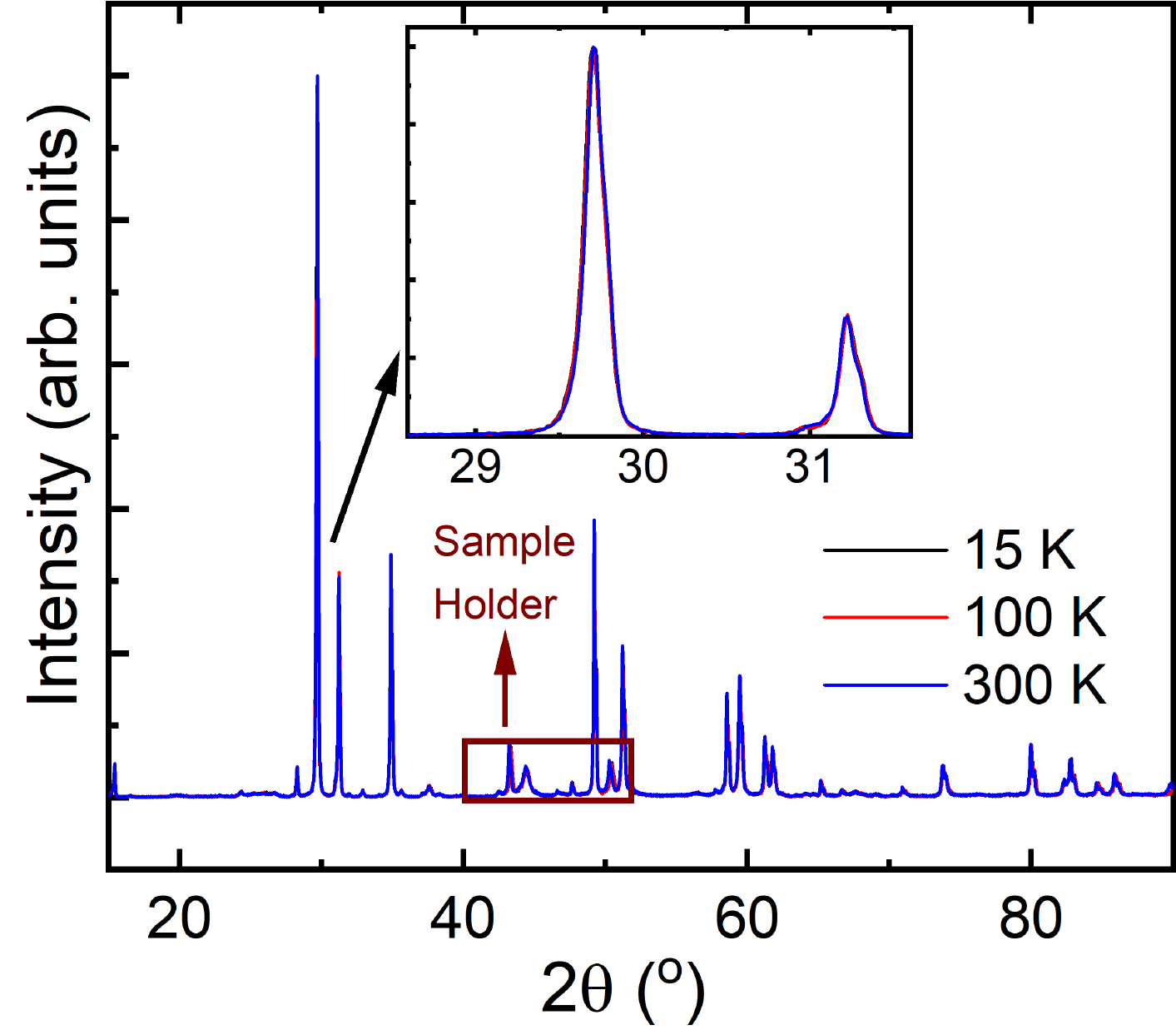}
		\caption{Powder XRD pattern for Y$_{3}$Cu$_{2}$Sb$_{3}$O$_{14}$ collected at 15, 100, and 300 K. The sample was placed on a stainless-steel holder.}
		\label{SF2}
	\end{center}
\end{figure}
\subsection{X-ray diffraction and crystal structure}
\begin{figure}[ht]
	\begin{center}
		\includegraphics[width=0.95\columnwidth]{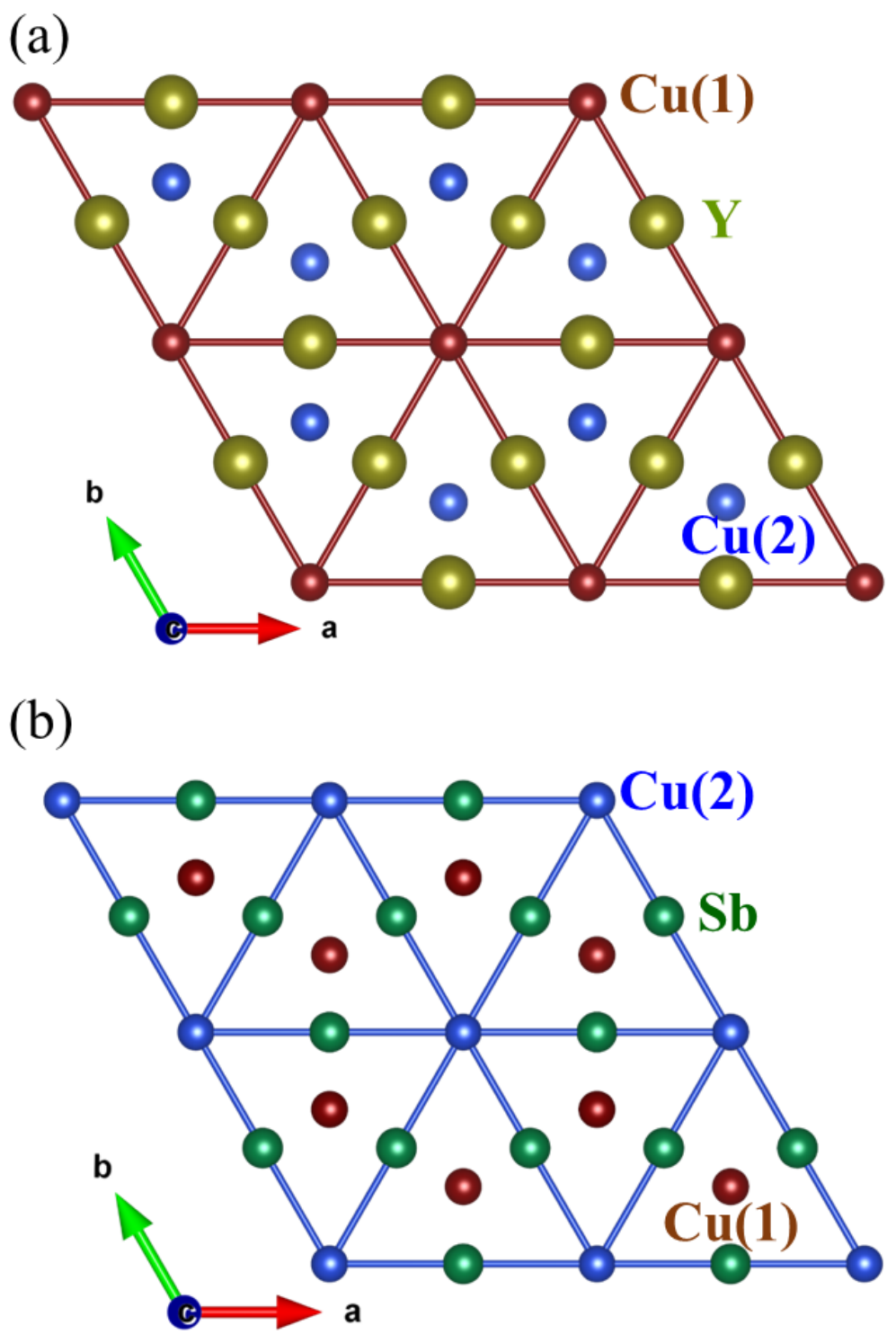}
		\caption{Triangular lattices formed by (a) Cu$^{2+}$(1) and (b) Cu$^{2+}$(2) ions in the $ab$ plane. The position of in-plane Y$^{3+}$ and Sb$^{5+}$ are also shown. Each Cu(2) site is centered in equilateral Cu(1) triangles in adjacent layers, and the same valid for Cu(1) sites.} 
		\label{SF3}
	\end{center}
\end{figure}
\begin{figure}[ht]
	\begin{center}
		\includegraphics[width=0.95\columnwidth]{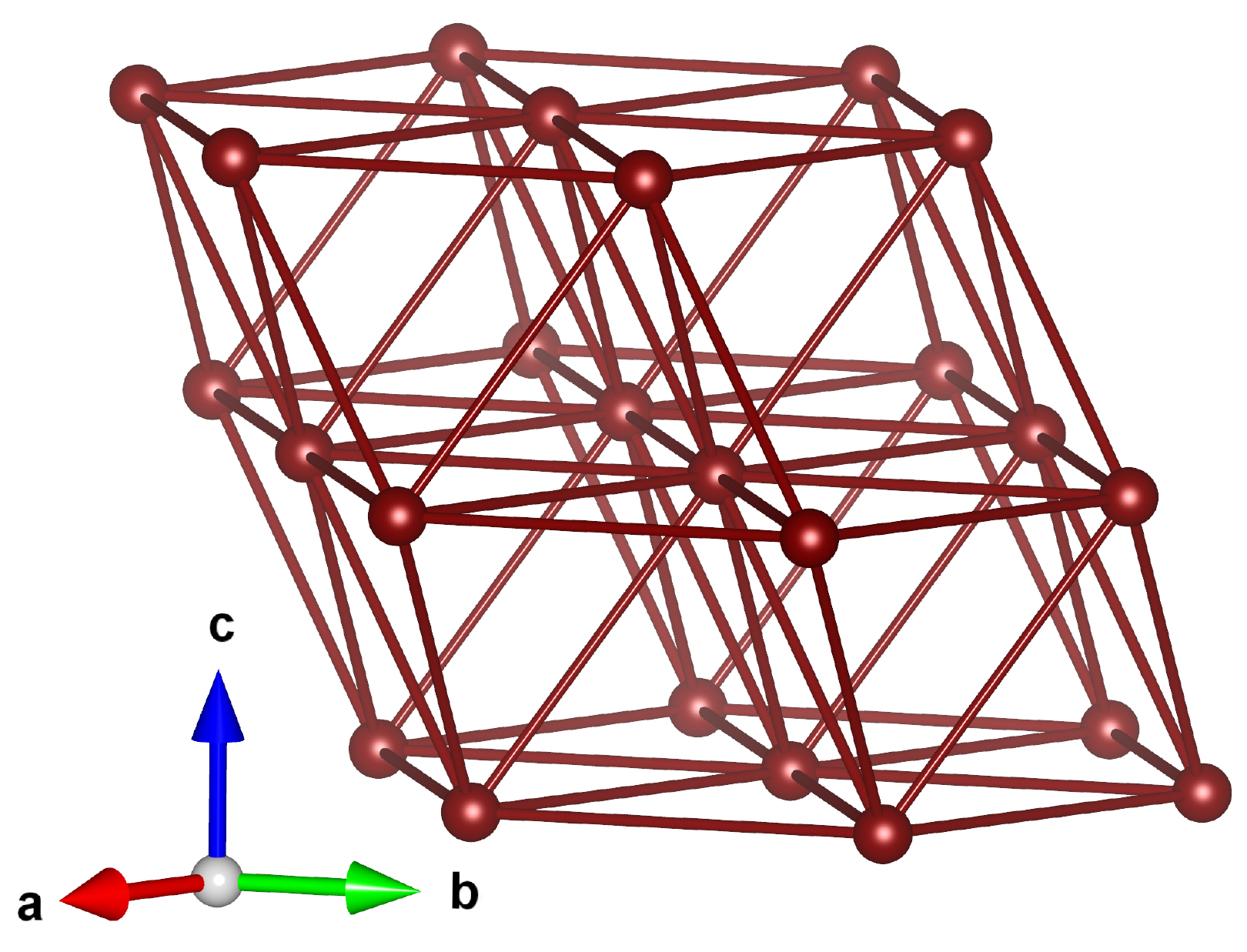}
		\caption{Frustrated 3D spin network of Cu$^{2+}(1)$. Also displayed are intraplane and interplane Cu(1)-Cu(1) linkages. The same pattern is also valid for Cu(2).} 
		\label{SF4}
	\end{center}
\end{figure}

Powder XRD pattern of Y$_{3}$Cu$_{2}$Sb$_{3}$O$_{14}$ measured at room temperature is shown in Fig. \ref{fig:S1}. There is no evidence of any secondary phase in XRD data. Rietveld analysis confirms that Y$_{3}$Cu$_{2}$Sb$_{3}$O$_{14}$ belongs to a rhombohedral structure with  $R\bar{3}m$ space group, with the Cu ions on sites $3a$ and $3b$ and the Y and Sb ions are on sites $9e$ and $9d$, respectively. We have also measured low-$T$ XRD down to 15 K as shown in Fig. \ref{SF2}. Refinement gives the same structure as at room temperature. The oxygens were initially placed in the positions found in Lu$_{3}$Mn$_{2}$Sb$_{3}$O$_{14}$ as a guide \cite{lee2023random}.  The results of the Rietveld refinements are summarized in Table \ref{tab:Rietveld-parameter_YCSO} and Table \ref{tab:Atomic-positions-in YCSO}. The obtained structural parameters for Y$_{3}$Cu$_{2}$Sb$_{3}$O$_{14}$ are in close agreement with the previous reports. It is also worthwhile to point out that in this rhombohedral structure, two Cu$^{2+}$ have different coordinations with O$^{2-}$ although they have the same oxidation state (2+). Cu(1) is in octahedral coordination (Cu(1)O$_6$) with Cu(1)-O(3) bond length of 1.946 \AA. Cu(2) is found in an eight-coordinated Cu(2)O$_8$ hexagonal bipyramids (trigonally-distorted cube) of two O(1) atoms along the local-[111] axis with Cu(2)-O(1) bond distances 1.82 \AA~ and a puckered six-membered ring of O(2) atoms with Cu(2)-O(2) lengths of 2.30 \AA. In this structure, the nearest Cu(1)-Cu(1) distance within the intralayer plane (7.40 \AA) is slightly greater than that between layers (7.14 \AA). The Cu(2)-Cu(2) bonds show a similar pattern though the nearest neighbor Cu(1)-Cu(2) bond length is 5.14 \AA. As shown in Fig. \ref{SF3}, when viewed along the $c$-axis,  Cu(1) ions form equilateral triangles in the $ab$-planes with Cu(2) ions (offset along the $c$-axis)  positioned at the centers of these triangles and vice versa. Likewise, Cu(1) ions in adjacent planes form a tetrahedral network (see Fig. \ref{SF4}). The Cu(1) and Cu(2) ions form two interpenetrating approximately FCC sublattices that realize a three-dimensional cubic network. Each of these sublattices (Cu(1) or Cu(2) alone) has the geometry of a nearly FCC lattice. It is well known that the FCC structure can be viewed as a stacking of two-dimensional triangular layers. The in-plane nearest-neighbour Cu(1)–Cu(1) (or Cu(2)–Cu(2)) bonds connect the sites of a 2D triangular lattice. These triangular layers are the natural units on which the frustrated in-plane antiferromagnetic couplings act.

Compared with the pyrochlore lattice ((RE)$_2$M$_2$O$_7$), the chemical formula can also be written as (Cu$_{0.25}$Y$_{0.75}$)$_2$(Cu$_{0.25}$Sb$_{0.75}$)$_2$O$_7$, where one-fourth of both Y$^{3+}$ and Sb$^{5+}$ ions are occupied by magnetic Cu$^{2+}$ ions in an ordered manner (Cu(1) and Cu(2) order to form Cu(1)Y$_3$ and Cu(2)Sb$_3$ planes, respectively, on both the RE and M pyrochlore sites). The ordering scheme results in the presence of two distinct 2D non-magnetic kagome layers (Y and Sb sites). This good kagome plane separation is likely due to the large ion size difference between Y$^{3+}$ and Cu$^{2+}$. It is notable that in the pyrochlore structure, the transition metal cations form the cubic (FCC) lattice with the stacking of the alternating (RE)$_3$M and M$_3$(RE) layers along the underlying four equivalent [111] FCC-directions, but in the rhombohedral structure, one of the four equivalent [111] corners becomes the trigonal $c$-axes. Examination of the X-ray diffraction patterns of Y$_{3}$Cu$_{2}$Sb$_{3}$O$_{14}$ reveals that the strongest peak (222) for pyrochlore at 2$\theta$ $\sim$ 30$^{\circ}$ has split into two peaks, qualitatively suggesting complete Cu–Y and Cu–Sb site-ordering~\cite{fu2014crystal, nandi2017}. 

\subsection{Magnetization}
\begin{figure*}
	\begin{center}
		\includegraphics[width=2.05\columnwidth]{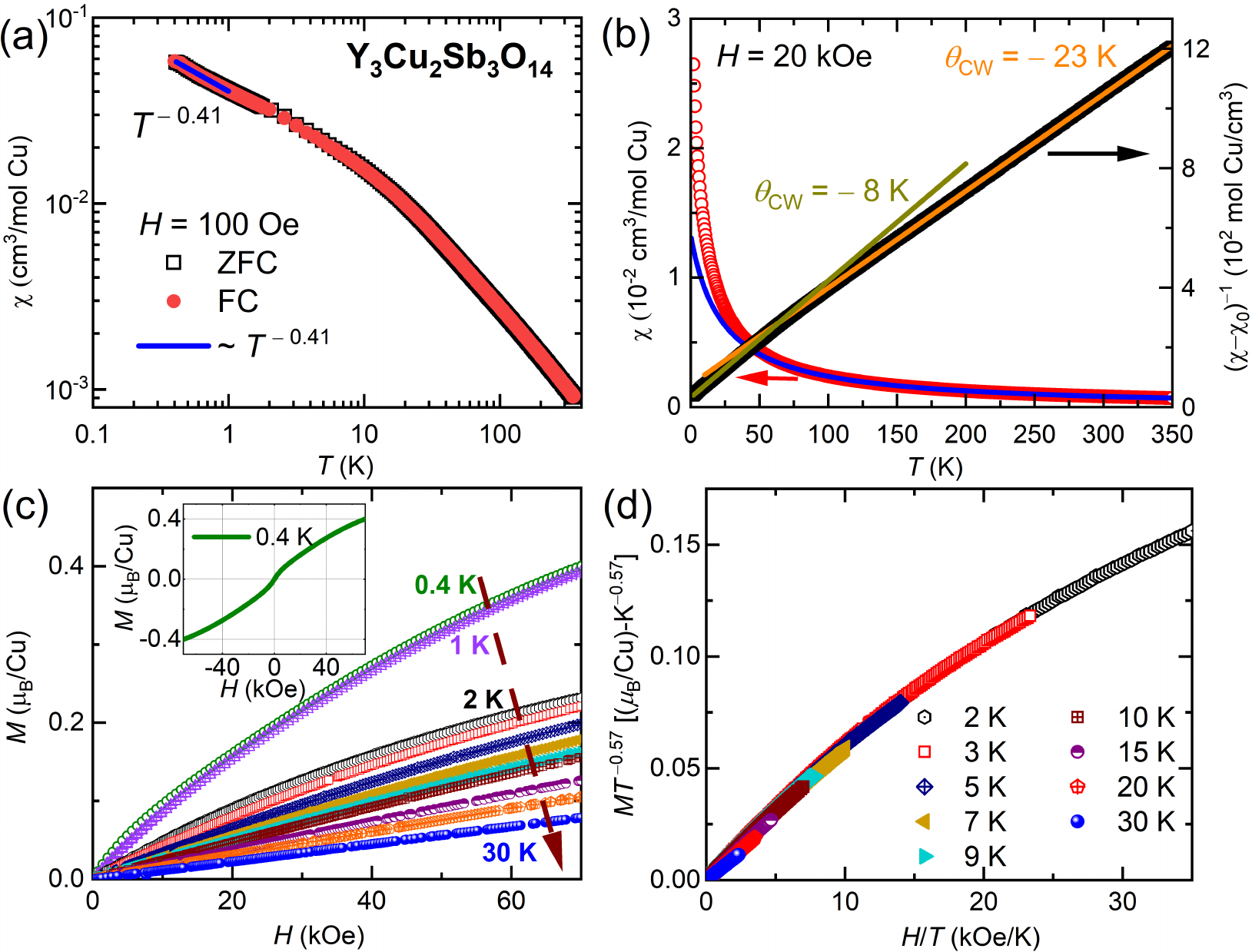}
		\caption{(a) Temperature dependence of the static magnetic susceptibility $\chi(T)$ of Y$_{3}$Cu$_{2}$Sb$_{3}$O$_{14}$ measured at $H=$ 100 Oe in zero field cooled (ZFC) (open symbols) and field cooled (FC) (closed symbols) processes on a log-log scale. No bifurcation between ZFC and FC $\chi(T)$ in is seen down to 0.40 K. The blue solid line depicts the low-temperature power-law fit ($\sim T^{-\alpha}$ with $\alpha = 0.41$). (b) Left $y$-axis: Temperature dependence of the static magnetic susceptibility $\chi(T)$ of Y$_{3}$Cu$_{2}$Sb$_{3}$O$_{14}$ measured at $H=$ 20 kOe. Right $y$-axis: The temperature dependence of inverse magnetic susceptibility data free from $\chi_{0}$. The Curie-Weiss fits are shown by the solid lines. (c) Field dependent isothermal magnetization ($M$ vs $H$) curve for the first quadrant at various temperatures. (inset) Isothermal magnetization($M$ vs $H$) curve at 0.4 K in the field range $-70$ kOe to 70 kOe. The absence of hysteresis in the $M$ vs $H$ plot indicates that there is no history dependence of magnetization in Y$_{3}$Cu$_{2}$Sb$_{3}$O$_{14}$. (d) Scaling of isothermal magnetization $M(H)$ with $MT^{-0.57}$ vs $H/T$.}
		\label{SF5}
	\end{center}
\end{figure*}

Temperature-dependent magnetic susceptibility ($\chi \equiv M/H$) measurements of Y$_{3}$Cu$_{2}$Sb$_{3}$O$_{14}$ measured in the temperature range of 0.40 K to  350 K with 100 Oe applied magnetic fields is shown in Fig. \ref{SF5}. The temperature-dependent dc susceptibility $\chi$ measured at $H= 20$ kOe, along with the  inverse susceptibility, is also shown in Fig. \ref{SF5}(b). No clear indication of any long-range magnetic ordering is observed down to 0.40 K and the data also do not show any anomaly in the dc and ac magnetic susceptibility data. Fig. \ref{SF5}(c) displays the magnetization curve at various temperatures. The zero-field-cooled (ZFC) and field-cooled (FC) data do not show any bifurcation in the dc susceptibility $\chi(T)$ which  rules out the occurrence of slow dynamics such as spin freezing behavior or glass-like state down to 0.40 K. This is further confirmed by the absence of hysteresis in the isothermal magnetization taken at 0.40 K (see Fig. \ref{SF5}(c) inset).

\begin{figure}[ht]
	\begin{center}
		\includegraphics[width=0.95\columnwidth]{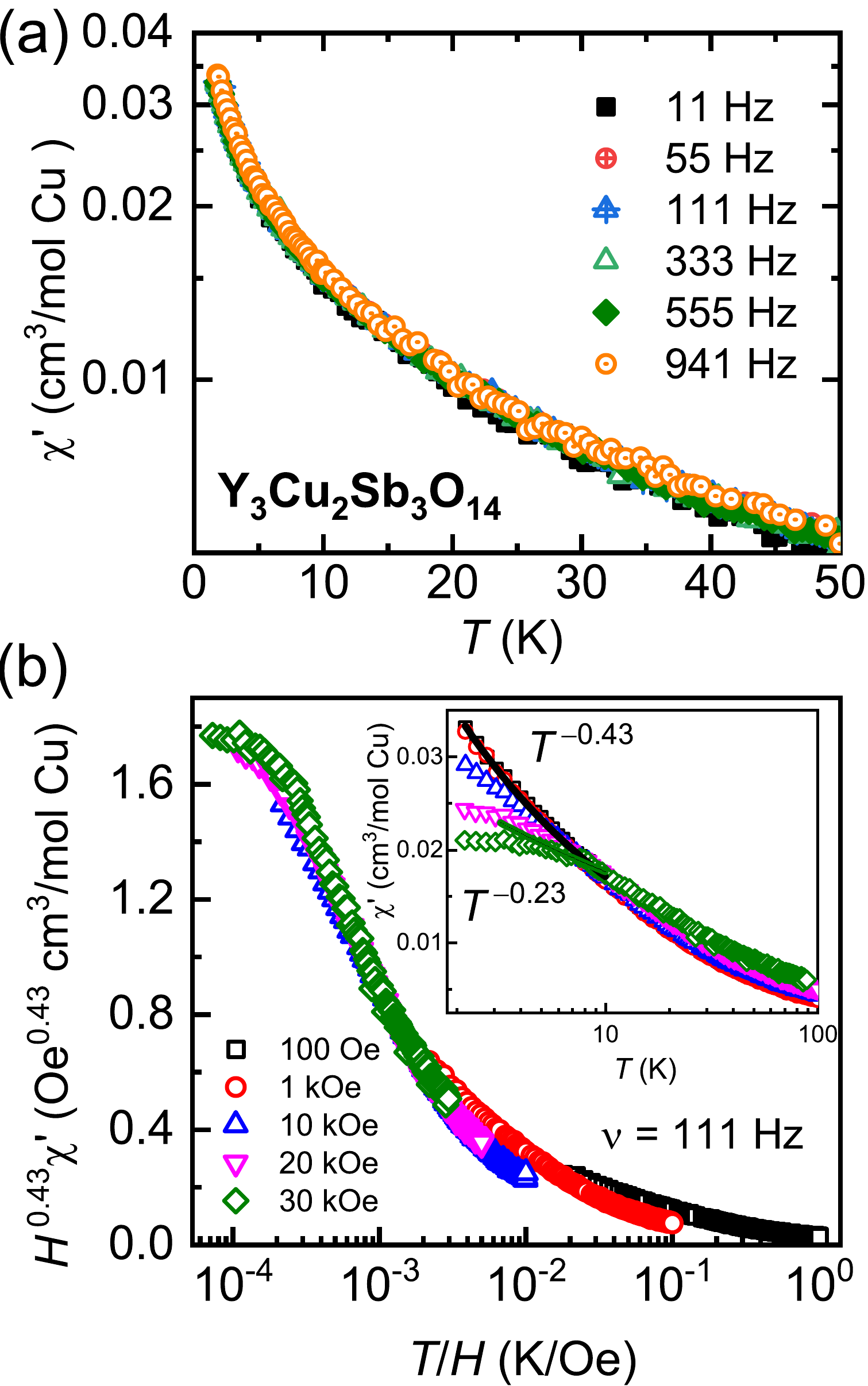}
		\caption{(a) Temperature dependence of the real component of the \textit{ac} susceptibility $\chi^\prime$, measured at various frequencies. No frequency-dependent behavior can
			be observed in the whole temperature range. (b) Scaling of the real component of the ac susceptibility $\chi^\prime(T)$ (at $\nu=$ 111 Hz) with $H^{0.43}\chi^\prime$ vs $T/H$. The inset plots $\chi^\prime(T)$ vs $T$ under various external dc fields. The solid lines are the power-law dependencies of $\chi^\prime$.}
		\label{SF6}
	\end{center}
\end{figure}

  The high-temperature magnetic susceptibility data for $H= 20$ kOe were fitted using the modified Curie-Weiss (CW) law $\chi$ = $\chi_{0}+C/(T-\theta_{\rm CW})$, in the temperature range of 150 K$\leq T\leq 350 $ K (see Fig. \ref{SF5}(b)). The high-temperature CW fit yields a negative $\theta_{\rm CW} = -23(1)$ K, which indicates antiferromagnetic interaction between the Cu$^{2+}$ spins.  The Curie-Weiss constant, $C$ turns out to be 0.31 cm$^3$ K/mol Cu, resulting in an effective moment $\mu_{\mathrm{eff}}$ = 1.57 $\mu_{B}$/Cu which is smaller than the free ion value and is thought to be due to covalency effects.  The obtained value of the temperature independent susceptibility ($\chi_{0}$) from the fit is $-1.37\times$10$^{-4}$ \,cm$^3$/mol Cu, which originates from core diamagnetism ($\chi_{core}$) as well as the van Vleck ($\chi_{vv}$) paramagnetic susceptibility. Adding the core diamagnetic susceptibility of the individual ions \cite{Bain}, we calculate the total $\chi_{core}=$ $-2.68\times$10$^{-4}$ \,cm$^3$/mol for Y$_{3}$Cu$_{2}$Sb$_{3}$O$_{14}$. The $\chi_{vv}$ was then estimated by subtracting $\chi_{core}$ from $\chi_{0}$, resulting in a negative van Vleck contribution, $\chi_{vv} = -0.03\times10^{-4}$ \,cm$^3$/mol Cu, which is unphysical and is perhaps due to our not reaching the asymptotic limit of the magnetic susceptibility at high-$T$ which brings some uncertainty to the inferred $\chi_{0}$. 
 
 We fitted $\chi(T)$ between 0.40 K - 1 K with a power-law,  $\chi(T) \propto T^{-\alpha}$ ($\alpha \thickapprox 0.41$) with no hint of the Curie tail for temperatures below 10 K (see Fig. \ref{SF5}(a)). Such a power-law law behavior points towards the presence of abundant low-energy excitations in a frustrated system. Fig. \ref{SF5}(d) shows a scaling of $MT^{-\gamma}$ as a function of $H/T$ at different temperatures. The data fall onto a single scaling curve with $\gamma=$ 0.57, which is close to the value $1-\alpha$.

 In addition, we probe the spin dynamics down to 2 K using ac susceptibility measurements. Fig. \ref{SF6}(a) displays the temperature dependence of the in-phase component of the ac susceptibility $\chi^\prime(T)$ at selected driving frequencies $\nu =$ 11, 55, 111, 333, 555, 941 Hz with 3 Oe ac drive field and no external dc field. $\chi^\prime(T)$ displays no discernible frequency dependence down to 2 K, indicating the absence of spin glass behavior. We present further supportive evidence for the dynamic scaling behavior by testing $H^{0.43}\chi^\prime$ as a function of $T/H$ (in Fig. \ref{SF6}(b)) for magnetic field up to 30 kOe. The scaling exponent ($\alpha=0.43$) is close to the value obtained from the power-law fit, ensuring that all data collapse onto a single curve at low temperatures. As the field increases,  $\chi^\prime$ ($H,T<$10 K) is systematically reduced (see inset of Fig. \ref{SF6}(b)) due to the freezing out of weakly coupled spins under an external magnetic field. However, it still maintains a power-law dependence while changing $\alpha$ from 0.43 at 100 Oe to 0.23 to 30 kOe. Similar behavior was observed in a certain class of frustrated systems \cite{choi2019exotic, lee2023random}.
  
  The commonly observed data collapse and power-law behavior ($\sim T^{-\alpha}$) in low-temperature  $\chi(T)$, $M(H)$, $\chi^\prime(T)$ and the magnetic specific heat $C_{mag}(T)$ data constitutes a hallmark of random singlet, spin liquid, and valence bond solid ground states with quenched disorders as reported in Cu$_2$IrO$_3$, A$_3$LiIr$_2$O$_6$ (A= H, Ag), ZnCu$_{3}$(OH)$_{6}$Cl$_{2}$, Lu$_3$Mn$_2$Sb$_3$O$_{14}$ \cite{choi2019exotic,kitagawa2018spin,bahrami2019thermodynamic,kenney2019coexistence, lee2023random}. Here in Y$_{3}$Cu$_{2}$Sb$_{3}$O$_{14}$, the presence of possible minor Cu(1)/Y and Cu(2)/Sb anti-site mixing might create disorder and give rise to  low-energy density-of-states $N(E)$, which follows a power-law behavior $N(E)\sim E^{-0.43}$~\cite{kimchi2018valence}. 

	\begin{figure*}
		\begin{center}
			\includegraphics[width=2.05\columnwidth]{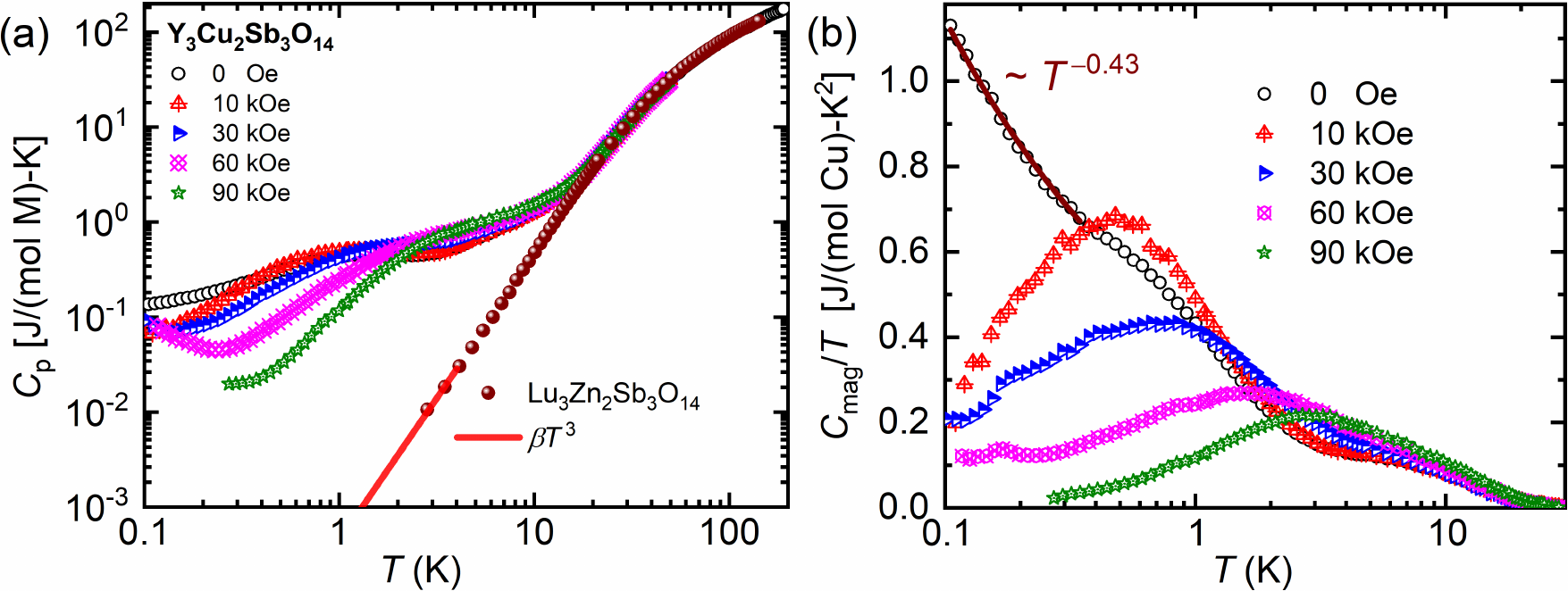}
			\caption{(a) Temperature dependence of the specific heat ($C_\mathrm{p}$) per mol ion (M$^{2+}$) at different magnetic fields up to 90 kOe on a log-log scale. The specific heat of zero field nonmagnetic Lu$_{3}$Zn$_{2}$Sb$_{3}$O$_{14}$ which represents the phonon contribution, is also shown for comparison. Red solid line is the fitted curve using $\beta T^3$ of the nonmagnetic specific heat data. (b) Magnetic specific heat divided by temperature $C_\mathrm{mag}/T$ vs $T$ for various magnetic fields. The low-$T$ power-law fit ($T^{-0.43}$) has been shown with a solid line at zero magnetic field.}
			\label{SF7}
		\end{center}
	\end{figure*}
\subsection{Specific heat}

\begin{figure}[ht]
	\begin{center}
		\includegraphics[width=0.95\columnwidth]{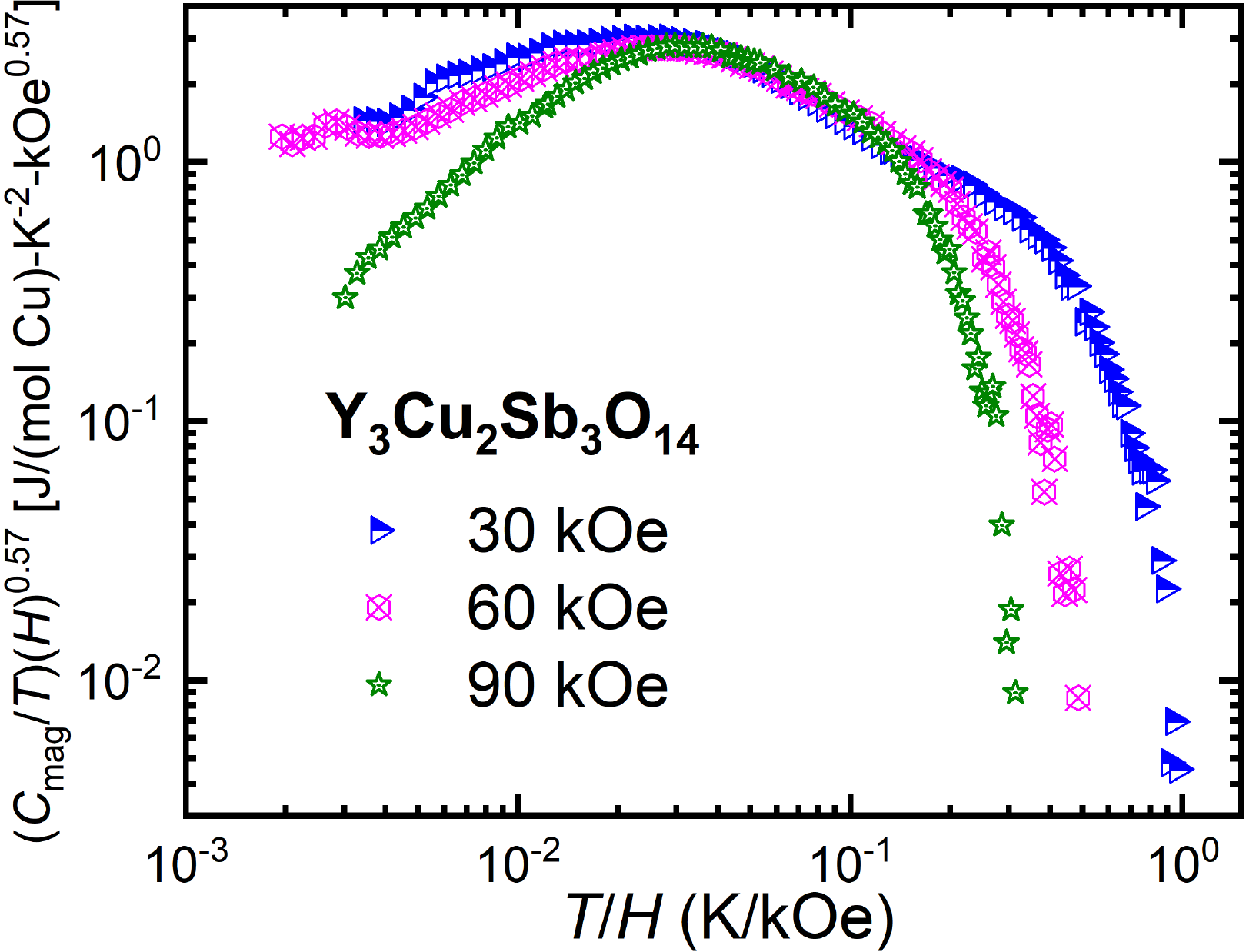}
		\caption{Scaling of magnetic specific heat $C_\mathrm{mag}$ on a log-log scale with $(C_\mathrm{mag}/T)(H)^{0.57} $ vs $T/H$.}
		\label{SF8}
	\end{center}
\end{figure}

 The phonon background of the specific heat $C_\mathrm{ph}(T)$ in the temperature range of 2 K to 150 K can be obtained by introducing a renormalization factor $b$ for the temperature of the specific heat data set $C^\mathrm{Zn}(T)$ of Lu$_{3}$Zn$_{2}$Sb$_{3}$O$_{14}$, i.e., $C_\mathrm{ph}(T) = C^\mathrm{Zn}(bT)$~\cite{bouvier1991specific}. 
 This was then extrapolated to lower temperatures using $C_\mathrm{ph}(T)= \beta T^3$ to analyse the low-$T$ data (see Fig. \ref{SF7}(a)). 
  Using a Debye-Einstein model with a combination of one Debye and three Einstein (1D+3E) functions yields a  similar lattice contribution. 
    In addition, the nuclear Schottky ($C_\mathrm{N}$) anomaly seen in high fields is also subtracted from the total specific heat $C_\mathrm{p}$ (not shown here). Fig. \ref{SF7}(b) shows a $C_\mathrm{mag} /T$ vs $T$ plot under the external fields of $H =$ 0$-$90 kOe. Despite a large $\lvert\theta_{\rm CW}\rvert$, $C_\mathrm{mag}$ shows no evidence for a transition to magnetic long-range order down to 0.10 K. The zero field $C_\mathrm{mag} /T$ vs $T$ data shows a power law behavior ($T^{-\alpha}$) with an exponent $\alpha=0.43$ at low temperatures, consistent with gapless magnetic excitations. 
  Also note that this exponent value is in agreement with the obtained $\alpha$ value from $\chi (T)$. We now examine the scaling of  $C_\mathrm{mag} (H, T)$ data using the relation, $(H)^\gamma C_\mathrm{mag}[H, T]/T = F_q [T/H]$ with a scaling function $F_q[T/H]$ \cite{kimchi2018scaling}. As shown in Fig. \ref{SF8}, we find data collapse of $C_\mathrm{mag}[H, T]/T$ in $T/H$ with $\gamma = 0.57 \thickapprox 1-\alpha$ in a limited range. This incomplete scaling of $C_\mathrm{mag} (T,H)$ over the full range might result from an inhomogeneous magnetic state with two different energy scales in Y$_{3}$Cu$_{2}$Sb$_{3}$O$_{14}$ \cite{lee2023random}. 

\subsection{Nuclear magnetic resonance}
\begin{figure}[ht]
	\begin{center}
		\includegraphics[width=0.97\columnwidth]{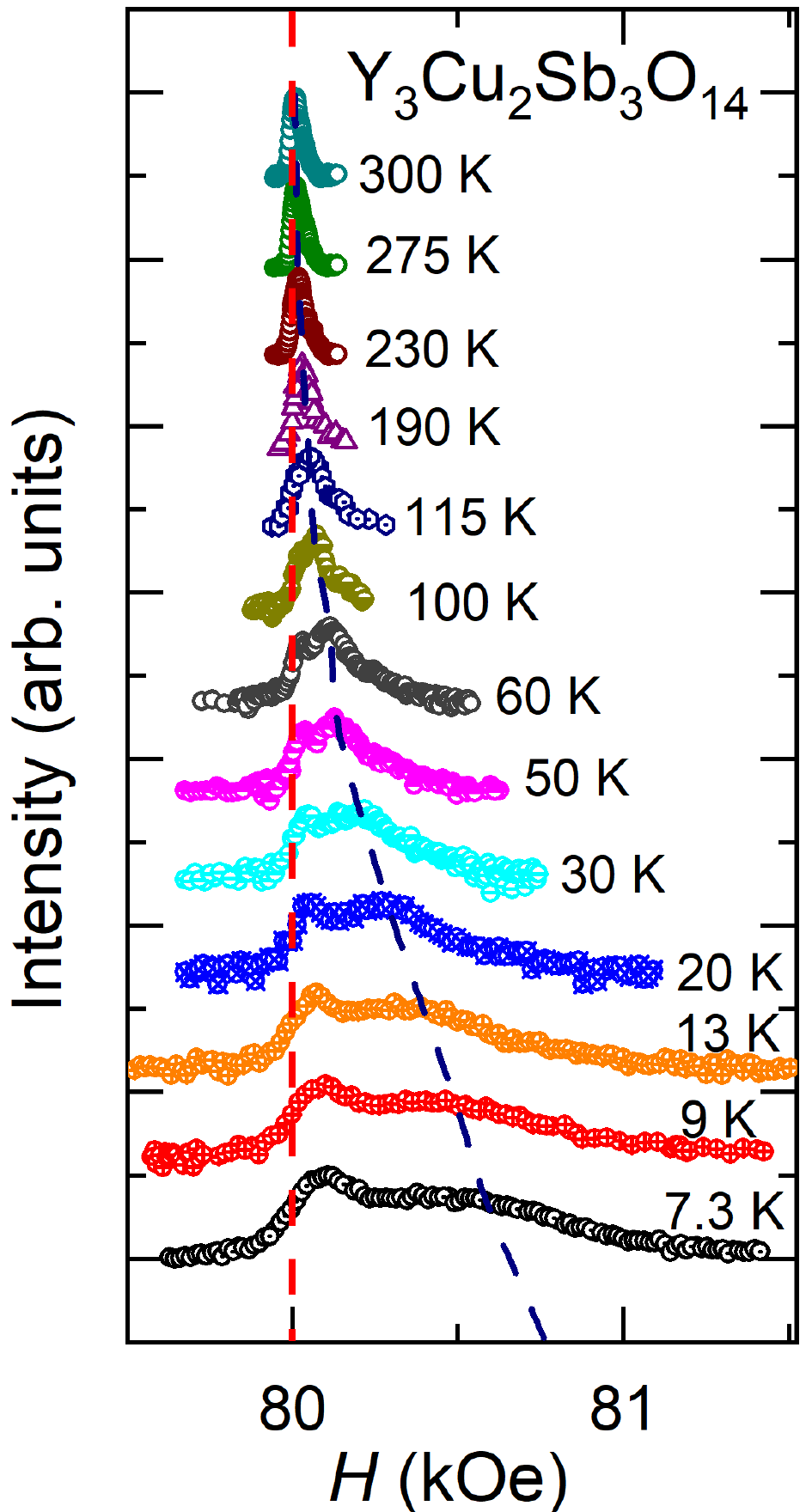}
		\caption {$^{89}$Y NMR field sweep spectra at various temperatures
			for Y$_{3}$Cu$_{2}$Sb$_{3}$O$_{14}$. The vertical red dashed line indicates the reference field (Y$_{3}$Zn$_{2}$Sb$_{3}$O$_{14}$). We have measured the NMR spin-lattice relaxation rate $1/T_1$ at each of the two peaks. The navy dashed line serves as a guide to the position of the right peak, which corresponds to the faster component.}
		\label{SF9}	
	\end{center}
\end{figure}

In order to develop a microscopic understanding of the magnetic properties, we performed $^{89}$Y (I = 1/2, $\gamma/2 \pi=$ 2.08583 MHz/T) nuclear magnetic resonance (NMR) measurements on polycrystalline Y$_{3}$Cu$_{2}$Sb$_{3}$O$_{14}$. This is the first NMR investigation of this family of compounds. 
The spin-echo intensity was obtained by integrating over the spin echo in the time domain. The final $^{89}$Y NMR spectrum was constructed by plotting the spin-echo intensity as a function of the applied magnetic field.
 Fig. \ref{SF9} displays the evolution of the $^{89}$Y NMR spectrum with temperature.
 To determine the $^{89}$Y NMR line shift as a function of temperature, we use a diamagnetic reference sample Y$_{3}$Zn$_{2}$Sb$_{3}$O$_{14}$ (measured at $T=$ 300 K).  

\begin{figure}[ht]
	\begin{center}
		\includegraphics[width=0.97\columnwidth]{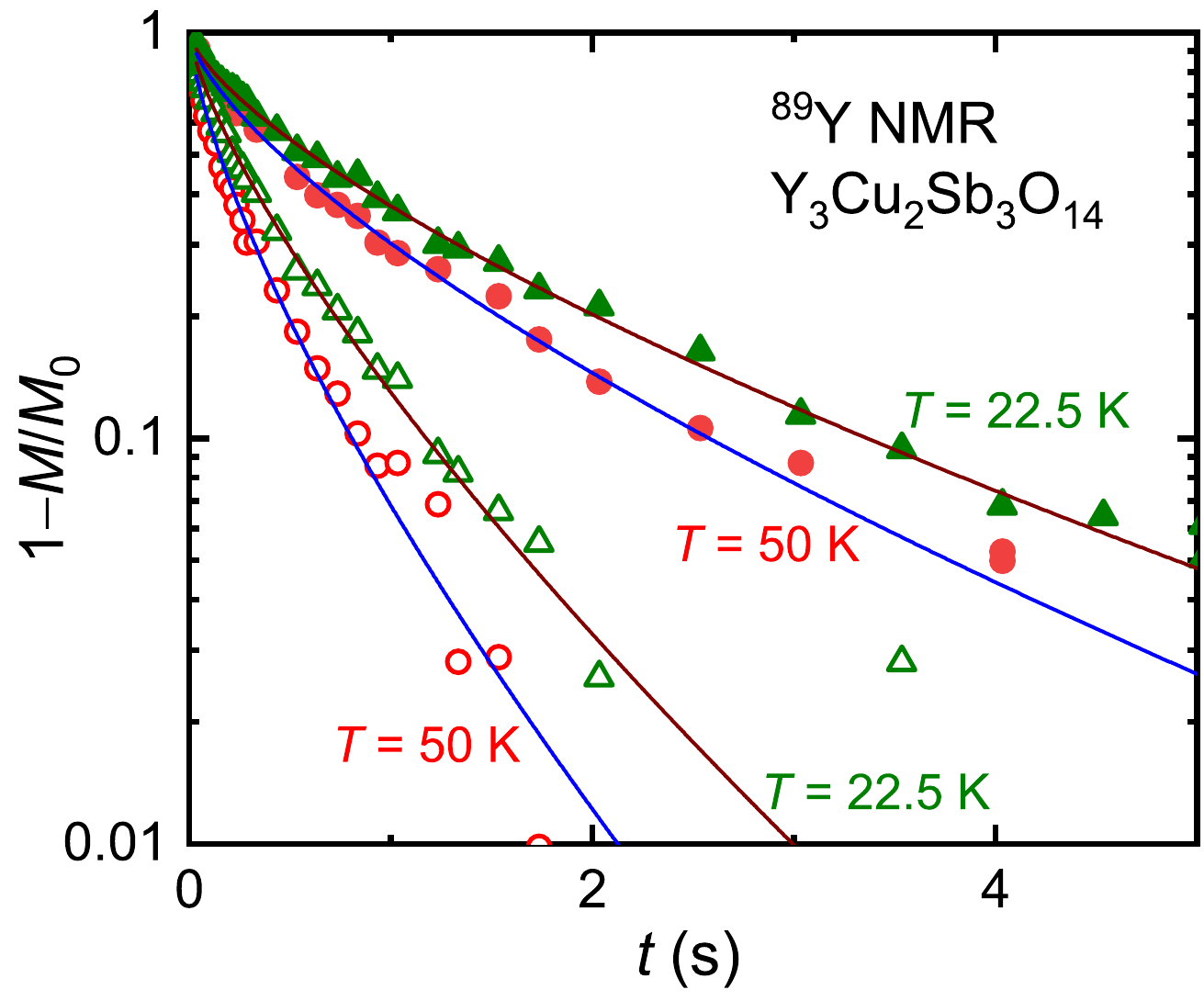}
		\caption {The recovery of the longitudinal nuclear magnetization of $^{89}$Y as a function of time delays $t$ is presented in semilog scale for two selected temperatures for representation. Open symbols represent the faster relaxation rate, while solid symbols represent the slower component.}
		\label{SF10}	
	\end{center}
\end{figure}
In Fig. \ref{SF10}, the recovery of the longitudinal nuclear magnetization $M(t)$ as a function of time delays $t$ is shown for two selected temperatures (22.50 K and 50 K). The intensity $M(t)$ is evaluated as the integral of Fourier-transformed spin-echo signal over the whole frequency region of the signal. $M(t)$ is well fitted with a stretched exponential function
{$M(t)=M_{0}[1-A\thinspace e^{-(t/T_{1})^{\beta}}]$}. Here $M_0$ is the saturation magnetization.
The decrease in stretching exponent $\beta$ is observed as temperature decreases. This results indicate inhomogeneous magnetic states which provide an additional relaxation mechanism. 

\begin{figure}[ht]
	\begin{center}
		\includegraphics[width=1.00\columnwidth]{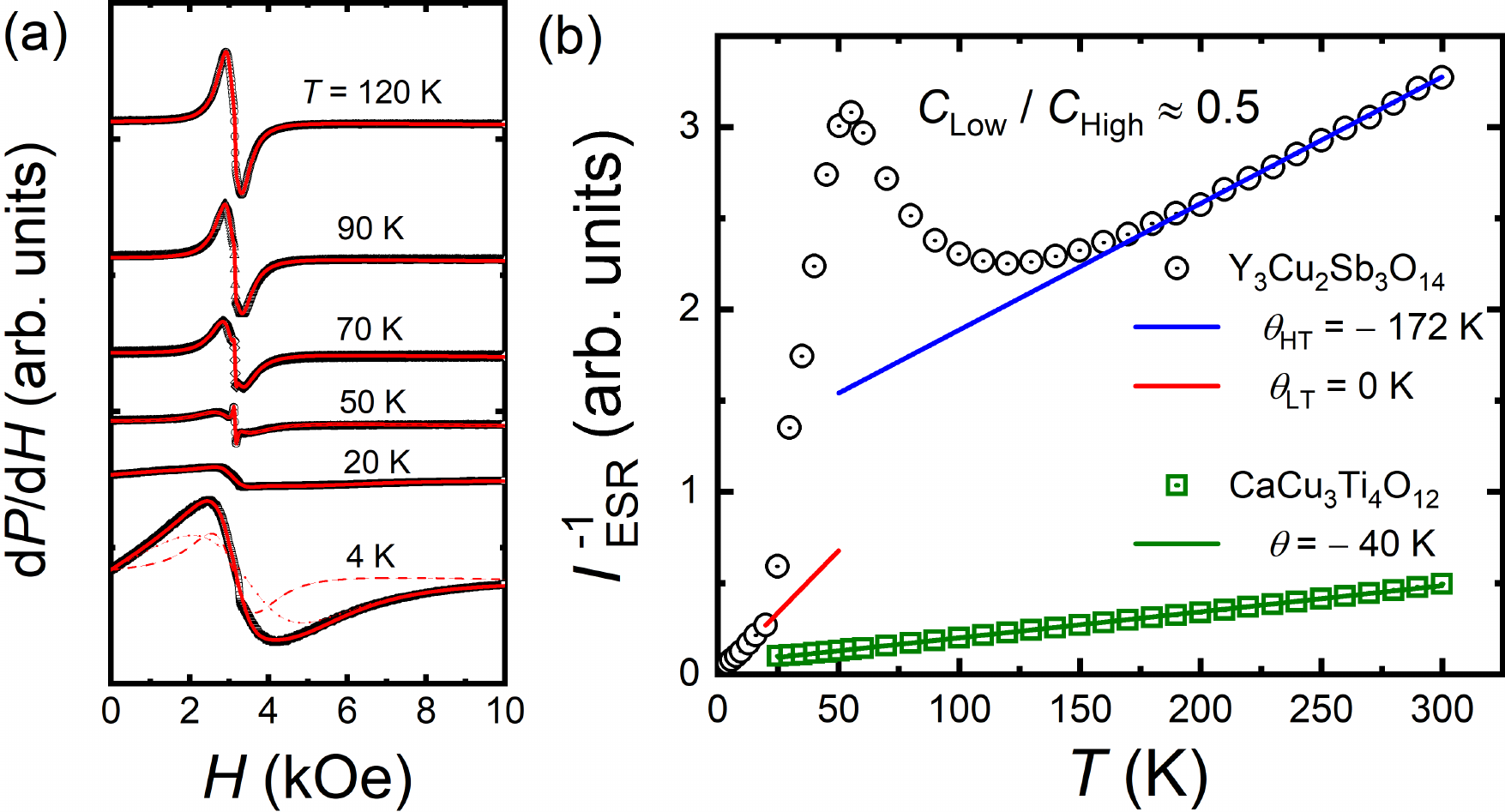}
		\caption {X-band ESR spectra (symbols) of Y$_3$Cu$_2$Sb$_3$O$_{14}$ at representative temperatures. Solid lines indicate fits by field derivatives of Lorentzian curves as described in the text. (b) Inverse integrated intensities of Y$_3$Cu$_2$Sb$_3$O$_{14}$ and CaCu$_3$Ti$_4$O$_{12}$, normalized by the number of copper ions per formula unit. Solid lines represent fits of the linear regimes by Curie-Weiss laws $I_{\rm ESR}^{-1} \propto (T-\theta)/C$ with CW temperatures $\theta_{\rm HT} = -172 (5)$\,K and $\theta_{\rm LT} = 0$\,K for high-$T$ and low-$T$ regimes of Y$_3$Cu$_2$Sb$_3$O$_{14}$ and $\theta = -40 (2)$\,K for CaCu$_3$Ti$_4$O$_{12}$, respectively.} 
		\label{SF11}	
	\end{center}
\end{figure}
\subsection{Electron spin resonance}

We performed temperature dependent X-band ESR measurements on polycrystalline samples of Y$_3$Cu$_2$Sb$_3$O$_{14}$. The ESR spectra (see Fig. \ref{SF11}(a)) contain a dominant exchange narrowed Lorentzian line above $T = 120$\,K due to fast electronic fluctuations of the Cu$^{2+}$ ions induced by exchange interaction. Its $g$-value of $g = 2.14(2)$ determined from the resonance field is typical for Cu$^{2+}$ ions in octahedral ligand field \cite{Abragam}. A second, about 10 times sharper but by two orders of magnitude less intense resonance coexists at $g = 2.13(1)$ presumably resulting from a small fraction of weakly coupled copper ions at grain surfaces or free spins around defects. On cooling through 120\,K down to about 50\,K, the dominant ESR line broadens, becomes weaker, and slightly shifts to lower resonance fields indicating the appearance of an internal field due to short-range correlations. On further cooling below 25\,K an additional very broad resonance line shows up with a similar $g$ value $g = 2.1(1)$, which finally dominates the ESR spectrum at 4 K.  
The temperature dependence of the inverse integrated ESR intensity $I_{\rm ESR}^{-1}$ follows different CW laws in the high-$T$ and low-$T$ limit with a transition regime 25 K $< T < 120$\,K (see Fig. \ref{SF11}(b)). From linear fits in the low- and high-$T$ limits, we find that the Curie constant at low-$T$ reduces to about one-half of the high-$T$ value. The corresponding CW-temperatures indicate strong antiferromagnetic interaction with $\theta_{\rm HT} = -172 (5)$\,K at high temperatures but effectively vanishing exchange with $\theta_{\rm LT} = 0$\,K at low temperatures.

In order to check, whether we are detecting all the copper spins, we calibrated the intensity with CaCu$_3$Ti$_4$O$_{12}$ where the ESR intensity is known to accurately represent the total spin susceptibility of the sample \cite{Zhuk}. As the slope is proportional to the inverse Curie constant and thus to the inverse number of spins, this indicates that above 120\,K only about 21(2)\% of the Cu spins contribute to the ESR signal in Y$_3$Cu$_2$Sb$_3$O$_{14}$ and drop to about one half of the high-$T$ value below 25\,K. This means that a large part of the copper spins is not detectable because of too fast relaxation resulting in a linewidth of several Tesla, which cannot be resolved in the typical field range of X-band experiments limited to a field range up to approximately 2 T (i.e. 20\,kOe). This is reminiscent of the high-$T_c$ cuprates which were found to be ESR-silent due to strong relaxation via spin-phonon interaction \cite{Kochelaev1999}. 

\begin{figure}
	\centering
	\includegraphics[width=1.00\columnwidth]{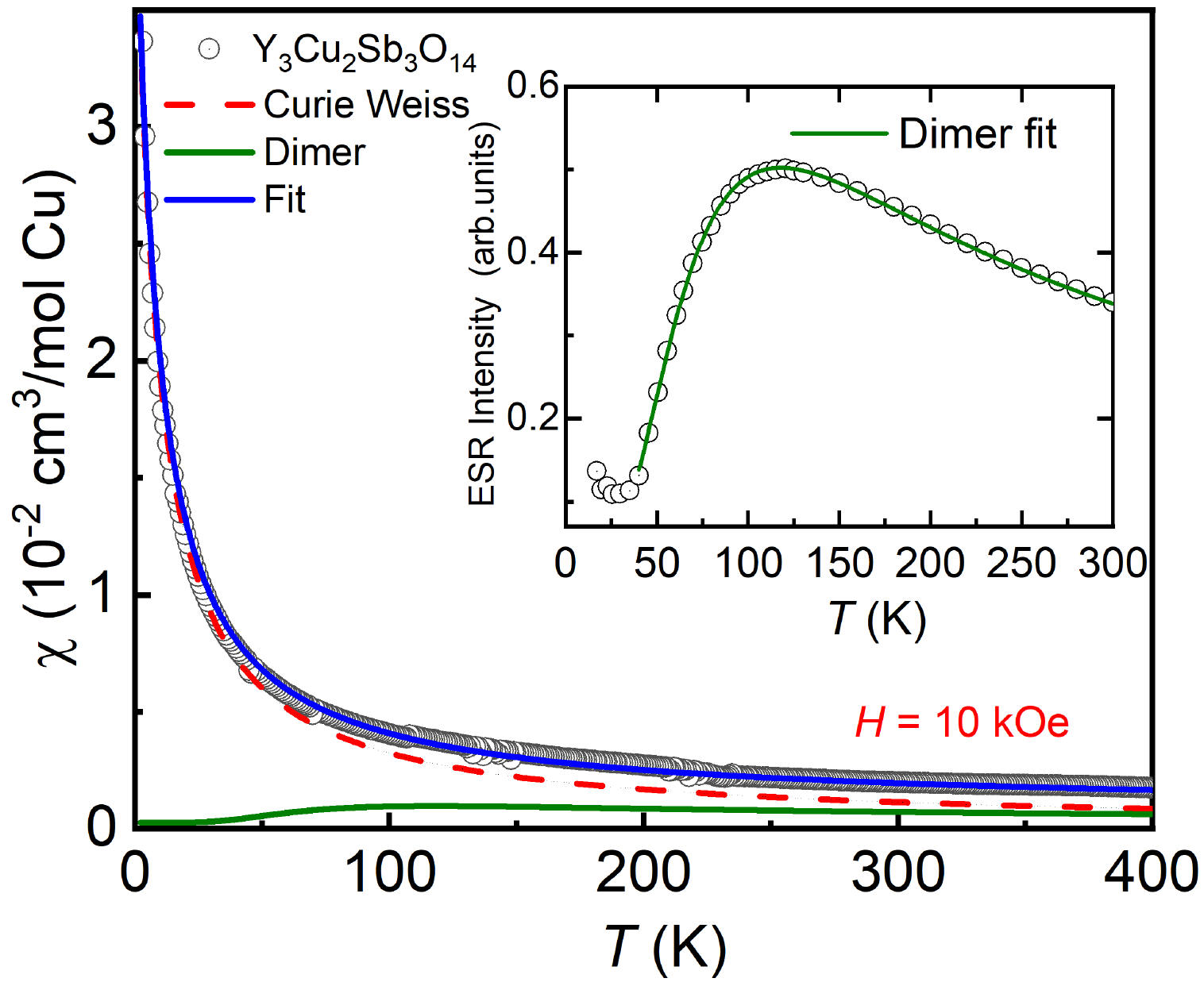}
	\caption{Temperature dependence of the magnetic susceptibility of Y$_3$Cu$_2$Sb$_3$O$_{14}$ measured at an external magnetic field of $H=10$\,kOe. The blue solid line indicates the fit of the data by the sum of a Curie-Weiss law (red dashed line) and Bleaney-Bowers dimer plus temperature independent van-Vleck term (green solid line). The ratio of Bleaney-Bowers contribution to Curie-Weiss contribution was fixed at $1:4$ at low temperatures. Inset: ESR intensity of the dominant Loremtz line above 25\,K. The green solid line represents the fit in terms of the Bleaney-Bowers formula (Eq. (\ref{eq1})) yielding $J/k_{\rm B} = -189 (5)$\,K.}
	\label{SF12}
\end{figure}

Further evaluation of the temperature dependent ESR intensity above 25\,K by means of a Bleaney-Bowers fit \cite{Bleaney}
\begin{equation} \label{eq1}
	I_{\rm ESR} \propto \frac{1}{T[3+\exp(-J/k_{\rm B}T)]}
\end{equation}
shown in the inset of Fig. \ref{SF12}, reveals the dimerization of the ESR active copper spins on cooling below 120\,K. This strongly supports the idea of formation of a VBS state. The intra-dimer exchange $J/k_{\rm B} = -189 (5)$\,K ($J \simeq -16.3$ meV) is large enough to warrant sufficient exchange narrowing making the ESR signal detectable. Using this information the static susceptibility can be described by the sum of a Curie-Weiss law ($\theta = -8.0 (5)$\,K), the Bleaney-Bowers dimer contribution, and a small temperature independent van-Vleck term of about $7.5 \times 10^{-4}$\,cm$^{3}$/mol Cu as illustrated in the main frame of Fig. \ref{SF12}. Note that the intra-dimer exchange constant determined from the fit of the ESR intensity, the $g$ value obtained from the resonance field and the ratio of dimerizing spins as compared to the residual spins were fixed at the values determined from our ESR experiments. We see that the fit turns out to be convincing. The derived CW-temperature agrees with the value obtained for the low-$T$ range in Fig. \ref{SF5}(b) and the underlying dimer contribution explains the enhanced CW temperature found in the high-$T$ range.  

To summarize the ESR measurements we can state that about 20\% of the Cu$^{2+}$ spins form strongly exchange coupled dimers, which pair as spin singlets below 120\,K, while 80\% remain unpaired but ESR silent due to fast relaxation. Note that regarding the NMR spectra in Fig. \ref{SF9}, the intensity ratio of the NMR lines corresponding to $^{89}$Y sites sensitive to Cu dimers and $^{89}$Y sites sensitive to unpaired Cu spins is also roughly $1:4$ at low temperatures. Only at temperatures below 25\,K a fraction of about 10\% of those unpaired spins becomes detectable by ESR. This fraction exhibits a zero CW temperature and, hence, is practically decoupled from the other spins with an average $\theta = -8$\,K determined from the susceptibility measurements.  

\subsection{Muon spin relaxation}

$\upmu\mathrm{SR}$ is an extremely sensitive local magnetic probe to detect small static internal fields ($\sim$ 0.1 Oe) arising from short-and long-range magnetic order or spin freezing. In order to determine the magnetic ground state and the nature of local spin dynamics, we employed the $\upmu\mathrm{SR}$ experiments on Y$_{3}$Cu$_{2}$Sb$_{3}$O$_{14}$. The zero-field $\upmu\mathrm{SR}$ spectra are well fitted by a Gaussian Kubo-Toyabe (KT) function, $G_{\mathrm{ZF}}^\mathrm{KT}$ multiplied by an exponential function, and a temperature-independent background, $P_\mathrm{BG}$ arising from the silver sample holder or the cryostat. $G_{\mathrm{ZF}}^\mathrm{KT} (\Delta, t)$ is the ZF  Kubo-Toyabe function reflecting the Gaussian distribution of randomly oriented or quasistatic local magnetic fields at the muon sites given by,
\begin{equation} \label{eq2}
	G_{\mathrm{ZF}}^\mathrm{KT} (\Delta, t) = \frac{2}{3} (1-\Delta^2t^2) \exp(-\frac{1}{2}\Delta^2t^2) + \frac{1}{3} ,
\end{equation}
\\
where $\Delta / \gamma_\mu$ describes the root-mean-square (rms) width of the Gaussian distribution and $\gamma_\mu = 2\pi \times 135.53$ MHz/T is the muon gyromagnetic ratio.

 \begin{figure}[ht]
 	\begin{center}
 		\includegraphics[width=0.95\columnwidth]{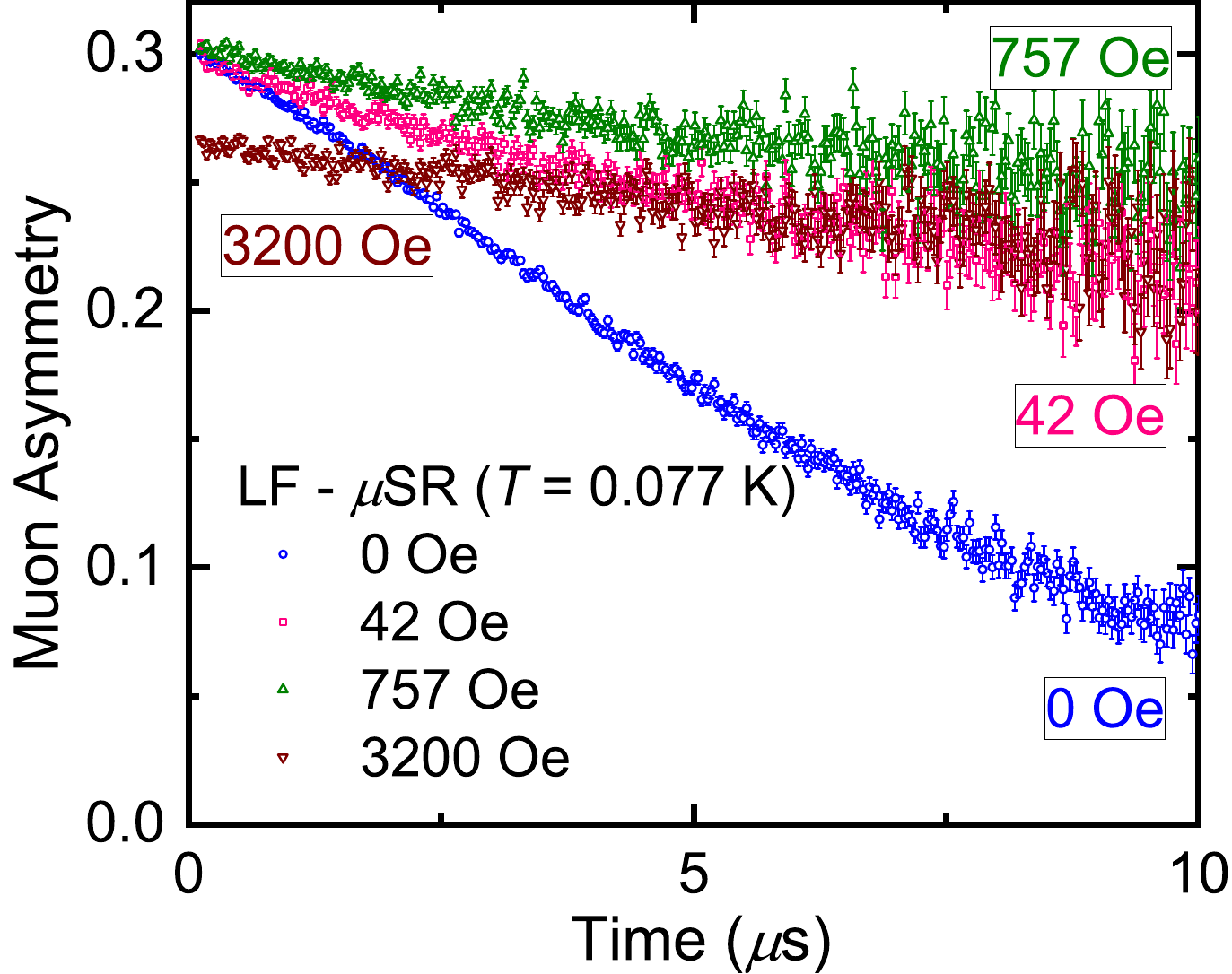}
 		\caption{Muon asymmetry as a function of decay time at 0.077 K for various longitudinal fields: 0 Oe, 42 Oe, 757 Oe, 3200 Oe. Muon asymmetry drop at 3200 Oe is due to the MUSR spectrometer limitations.}
 		\label{SF13}
 	\end{center}
 \end{figure}

At low temperature ($T< 1$ K), the $\lambda_{\mathrm{ZF}}$ value is roughly same order magnitude with an average value of $\Delta$.
The parameter $\Delta \thickapprox$  0.1 MHz corresponds to an rms quasistatic local field $B_{loc} \thickapprox$ 1.17 Oe, therefore, decoupling of muons is expected with an applied longitudinal field (LF) of 12 Oe ($\sim$ 10$B_{loc}$). 

It is interesting to note that the onset of plateau ($T \thickapprox 10$ K) feature corresponds to roughly the same temperature as the broad hump seen in the magnetic specific heat, similar to that other QSL compounds  \cite{Pula2024, bhattacharya2024evidence}.   

Muon decoupling experiments were performed in several applied longitudinal fields (LF) at 0.077 K are shown in Fig. \ref{SF13}. In the presence of static magnetism, an LF of 10$B_{loc}$ could completely suppress the static field and decouple the muon spins from the influence of that internal static fields. Decoupling is not possible even in an LF of 3200 Oe, which is about 2730 Oe times larger than $B_{loc}$, clearly reveal that the magnetic ground state of Y$_{3}$Cu$_{2}$Sb$_{3}$O$_{14}$ is entirely dynamic in nature, as would be expected for a quantum spin liquid.   
\subsection{Electronic structure calculation}

\begin{figure}[ht]
	\begin{center}
		\includegraphics[width=0.95\columnwidth]{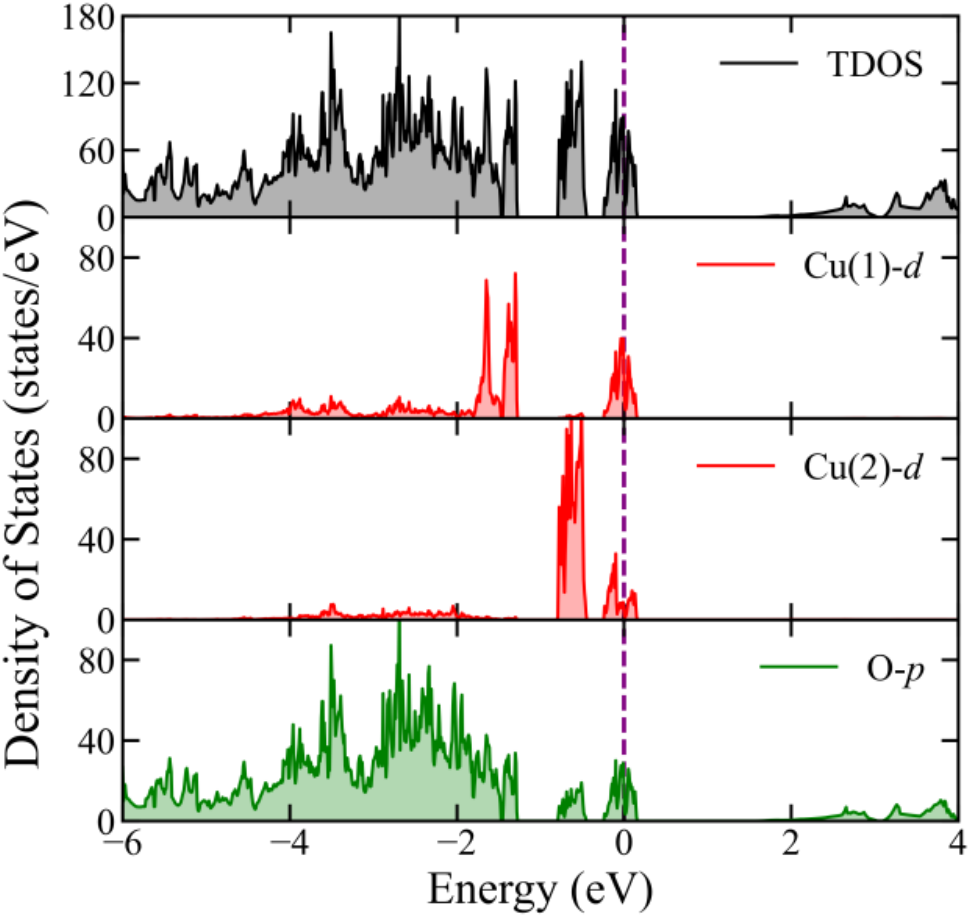}
		\caption{Non-spin polarized total (grey), Cu-$d$ partial (red) and O-$p$ partial (green) DOS for Y$_3$Cu$_2$Sb$_3$O$_{14}$.}
		\label{SF14}
	\end{center}
\end{figure}

The electronic structure calculations were done for an ordered crystal structure, such that within the unit cell of Y$_3$Cu$_2$Sb$_3$O$_{14}$, both Cu(1) and Cu(2) atoms form triangular network. To construct such a crystal structure we have considered an $2{\times}2{\times}1$ super-cell (containing 264 atoms) of the three formula unit primitive unit cell of the compound Y$_3$Cu$_2$Sb$_3$O$_{14}$. The distances between the six nearest neighbors in the triangular network are equal (7.40 \r{A}) for both Cu(1) and Cu(2) atoms. Here Cu(1) is in octahedral environment and Cu(2) is in hexagonal  bipyramidal environment. The local Cu(1)O$_6$ octahedral environment breaks the Cu(1)-$d$ orbitals into triply degenerate $t_{2g}$ and doubly degenerate $e_g$ states. The Cu(2)-$d$ orbitals are split into doubly degenerate e$_{1g}$, doubly degenerate $e_{2g}$ and non-degenerate $a_{1g}$ states in presence of the Cu(2)O$_{8}$ hexagonal bipyramidal environment. Our theoretically calculated non spin-polarized total and partial density of states plot is shown below (Fig. \ref{SF14}).

In the crystal structure of Y$_3$Cu$_2$Sb$_3$O$_{14}$ the distorted octahedra around Cu(1) and the distorted hexagonal bipyramid around Cu(2) do not share common oxygen atoms. As a consequence, the antiferromagnetic exchange coupling proceeds via super-super exchange path Cu(1)-O$\mathrm{\cdots}$O-Cu(2) rather than direct Cu(1)-O-Cu(2) bridges. The nearest-neighbor Cu(1)-O$\mathrm{\cdots}$O-Cu(2) links have a bent geometry with $\angle$Cu(1)-O$\mathrm{\cdots}$O-Cu(2) $\approx$ $90^{\mathrm{o}}$, which suppresses the O$_{p_\sigma}$-O$_{p_{\sigma}}$ overlaps between the two oxygen atoms mediating the super-super exchange and yields a very small exchange coupling. In contrast, for the body diagonal path, the $\angle$Cu(1)-O$\mathrm{\cdots}$O-Cu(2) is nearly linear ($\approx$ $180^{\mathrm{o}}$), maximizing the O$_{p_\sigma}$-O$_{p_\sigma}$ overlap resulting in a larger value of the exchange coupling. This was further corroborated by calculating the effective hoppings between the $d$-orbitals of Cu(1) and Cu(2) by downfolding the O, Y and Sb states, where the nearest neighbor hopping $t_\mathrm{AB}$ was found to be substantially weaker in comparison to the hoppings along the body and face diagonal. 

\bibliography{ref}